\documentclass[a4paper,11pt]{article}
\usepackage{jheppub} 
\usepackage{tcolorbox}
\usepackage{xcolor}
\usepackage{lipsum}
\usepackage{tikz}
\usepackage[compat=1.1.0]{tikz-feynman}
\usepackage{graphicx}
\usepackage{subfig}
\usepackage{color}

\arxivnumber{2304.08425} 

\title{\boldmath Two-loop Vilkovisky-DeWitt effective action for scalar field minimally coupled with gravity}

\author[a]{Ketankumar Jadav,}
\author[a]{Sukanta Panda,}
\author[a]{and Abbas Altafhussain Tinwala}
\affiliation[a]{Department of Physics, Indian Institute of Science Education and Research Bhopal - 462066, India}

\emailAdd{jadav18@iiserb.ac.in}
\emailAdd{sukanta@iiserb.ac.in}
\emailAdd{abbas18@iiserb.ac.in}

\abstract{It is known that Vilkovisky-DeWitt's effective action is gauge invariant off-shell and independent of quantum field parametrization. In this work, we perform the calculation of the divergences in two-loop Vilkovisky-DeWitt's effective action for the theory of a massive scalar field minimally coupled to gravity. We quantize both gravity and scalar field around a flat Minkowski metric and a constant scalar field background. The quantum corrections are computed up to second order in the background scalar field for the purpose of mass renormalization.}

\begin{document}
\maketitle
\flushbottom

\section{Introduction}
   Previous studies have shown that the standard effective action for gauge theories suffers from off-shell gauge dependence and the choice of parametrization for quantum fields. The unique effective action formalism proposed by Vilkovisky and DeWitt\cite{unique,lavrov,odin, tomsbook} resolves this issue by yielding an effective action completely free from the gauge and parametrization ambiguity.

   \noindent
   There has been extensive work on Vilkovisky-DeWitt's (VD) effective action at one loop level \cite{odintsov01, odintsov02}.
   Recent works include \cite{giacchini2}, which focuses entirely on the issue of gauge dependence arising due to the metric in the configuration space itself depending on the gauge fixing parameter, and \cite{giacchini1} devoted to the subject of the renormalization group in effective quantum gravity. Calculations of effective action using standard formalism beyond one-loop have been performed in the past for various theories \cite{odin2loop, 2loopfermion}. However, according to our knowledge, an explicit calculation of the VD effective action beyond one-loop is still lacking. In this paper, we present a thorough calculation of the VD effective action up to two-loop for the following theory,
   
   \begin{align}\label{1}
       S = \int d^4x \sqrt{|g(x)|}\left(-\frac{R}{2\kappa^2}+\frac{1}{2}\partial_\mu\phi\partial^\mu\phi+\frac{1}{2}m^2\phi^2\right),
   \end{align}

\noindent
where $\kappa^2 = 32\pi G$.
For any theory involving gravity, it is hard to calculate the standard effective action beyond one-loop, let alone VD effective action, which is why we have chosen to work with the most simple form of action possible involving gravity and a massive scalar field. Beyond one-loop, the simple functional form, even in the case of standard effective action, is lost. As a result, the sophisticated heat kernel technique, as well as its extensions to cover a wide range of nonminimal operators known to give closed-form results for loop corrections inapplicable beyond one-loop \cite{nonminimal}. In fact, there is no known direct way of obtaining a  two-loop effective action, in the light of which we resort to the most commonly used method for the evaluation of VD effective action, which involves perturbative expansion in orders of background fields, followed by calculating the loop integrals, as outlined in \cite{mackaytoms, aashish2021covariant,sandeep,ourpaper2,genback}. We focus on the two-loop mass renormalization for which evaluating the effective action up to the second order in the background scalar field is enough. We choose a flat background metric and a constant scalar field background for this work to ease our calculations. The paper is organized as follows. We present the expression for two-loop VD effective action in Sec \ref{SecII} with the derivation placed conveniently in Appendix \ref{AppA}. We then proceed with the explicit calculation of the two-loop Vilkovisky DeWitt effective action for the theory \eqref{1} in Sec \ref{SecIII}. We conclude the work in Sec \ref{SecIV} along with a discussion on mass renormalization.
    
\section{Effective Action}\label{SecII}
Before we present the expression for two-loop VD effective action, we familiarise ourselves with the DeWitt notation, which we will be following in this paper. The symbols used to represent the properties of a field are labeled as $\varphi^i$, where the discrete field index and the field's spacetime argument are condensed into the single label $i$. For instance, if the field is a scalar field, then $\varphi^i$ is equivalent to $\phi(x)$, and if it is a vector field, then $\varphi^i$ is equivalent to $A_\mu(x)$, while for a second-rank tensor field, $\varphi^i$ is equivalent to $g_{\mu\nu}(x)$. Additionally, the following summation convention is used in $n$ dimensions:

\begin{equation}\label{2}
    \varphi^i B_{ij} \varphi^j = \int d^nx\int d^nx' \varphi^I(x) B_{IJ}(x,x') \varphi^J(x'),
\end{equation}

\noindent
where capital Latin letters $(I, J,...)$ are used as a placeholder for conventional field indices,

\noindent
Also from,

\begin{align}\label{3}
     \varphi^I(x) &= \int d^nx' |g(x')|^{1/2}\delta(x,x') \varphi^I(x')\nonumber\\
     & = \int d^nx' \tilde\delta(x,x') \varphi^I(x').
\end{align}

\noindent
 We define $\tilde\delta(x,x')$:
 
\begin{equation}\label{4}
    \tilde\delta(x,x') = |g(x')|^{1/2}\delta(x,x'),
\end{equation}

\noindent
where $\delta(x,x')$ is the conventional bi-scalar Dirac $\delta$-distribution.

\noindent
Having established a proper language, we begin with the recursion relation for the standard effective action in the path integral formulation for a general theory of $N$-bosonic fields,

\begin{equation}\label{5}
	    	\exp{\frac{i}{\hbar} \Gamma [\bar{\varphi}]} = \int [\mathcal{D} \varphi] \exp{\frac{i}{\hbar}\left\{S[\varphi] -(\varphi^i- \bar{\varphi}^i)\frac{\delta\Gamma[\bar{\varphi}]}{\delta \bar{\varphi}^i}\right\}},
\end{equation}

\noindent
where, $[\mathcal{D} \varphi]$ indicates integral to be taken over all field configurations and $\bar\varphi^i$ is the background field. 

\noindent
For VD effective action, we have the following expression
(We intend not to present the intricate details of steps used to arrive at the relation since an extensive amount of work has been devoted to it. We suggest readers the standard texts \cite{tomsbook,Odintsovbook} for a detailed review),

\begin{align}\label{6}
         \exp{\frac{i}{\hbar} \Gamma[\bar{\varphi};\varphi_*]} &= \int \prod_id \varphi^i |g[\varphi]|^\frac{1}{2} 
        \left|\Delta[\varphi_*;\varphi]\right| 
         \exp{\frac{i}{\hbar} \left\{ S[\varphi] +  \frac{\delta\Gamma[\bar{\varphi};\varphi_*]}{\delta\sigma^i[\varphi_*;\bar{\varphi}]}  \left( \sigma^i[\varphi_*;\bar{\varphi}] -  \sigma^i[\varphi_*;\varphi] \right) \right\}}.
\end{align}

\noindent
We observe that Synge's world function $\sigma^i$, given by,

\begin{equation}\label{7}
    \sigma^i[\varphi_*;\varphi] = g^{ij}[\varphi_*]\frac{\delta}{\delta \varphi_*^j}\sigma[\varphi_*;\varphi],
\end{equation}

\noindent
replaces the coordinate difference $\varphi^i-\bar\varphi^i$ in the standard effective action \eqref{5}, which addresses the issue of non-covariance. The quantity $g_{ij}[\varphi]$ is the metric over the field configuration space at field coordinate $\varphi^i$. The measure $\Delta[\varphi_*;\varphi]$, called the  Van Vleck-Morette determinant, is a by-product that appears when we switch from the field variable $\sigma^i$ to $\varphi^i$ in the path integral. The choice of field coordinates $\varphi_*$ here is arbitrary, much like that for gauge parameters in gauge theories. DeWitt's choice $\varphi_*=\bar\varphi$ turns out to be the easiest for computational purposes. However, we leave this replacement aside until having expanded the VD effective action in loop order.

\noindent
In the case of gauge theories, we expect effective action to be gauge invariant as well as independent of gauge condition.
For gauge theories, it becomes mandatory that off-shell effective action be independent of field parametrization. A simple way to see this is to note the fact that when one employs the Fadeev-Popov procedure to convert a gauge action
to a non-gauge one by choosing a specific gauge slice, then geometrically, this means we are selecting a specific field
parametrization. Thus if effective action depends on field parametrization, then it will also depend on what gauge
choice we make. 

\noindent
For gauge theories, the action functional remains invariant under transformations belonging
to the gauge group. The gauge transformations may be written as,

\begin{equation}\label{8}
    \delta \varphi^i = K^i_\alpha[\varphi] \ \delta \epsilon^\alpha,
\end{equation}

\noindent
where $\epsilon^\alpha$ is a set of parameters that characterize the transformation and $K^i_\alpha[\phi]$ is the component of the generator of the transformation, $\textbf{K}_\alpha[\phi] = K^i_\alpha[\phi] \dfrac{\delta}{\delta \phi^i}$. Here $K^i_\alpha$ forms a Lie Algebra.

\begin{equation}\label{9}
   \bigl[ \textbf{K}_\alpha[\phi], \textbf{K}_\beta[\phi] \bigr] = - f_{\alpha \beta}^\gamma[\phi] \textbf{K}_\gamma[\phi],
\end{equation}

\noindent
with $f_{\alpha \beta}^\gamma[\phi]$ as the structure constant for the Lie Algebra. The recursion formula for the VD effective action reads,

\begin{align}\label{10}
      \text{exp} \Bigg(i \dfrac{\Gamma [\Bar{\varphi};\varphi_*]}{\hbar}\Bigg) &= \lim_{\alpha \rightarrow 0} \int \prod_{i} d \sigma^i \sqrt{g[\varphi^*]} [Dc][D \Bar{c}] \ \text{exp} \Bigg(\dfrac{i}{\hbar} \Big(S[\varphi] + S_{GF}[\varphi,\alpha] + S_{GH}[c,\Bar{c}] + \nonumber \\& (v^i-\sigma^i)\dfrac{\delta \Gamma[v;\varphi_*]}{\delta v^i}\Big)\Bigg).
\end{align}

\noindent
Just like for standard effective action for gauge theories, we also have here the gauge fixing $S_\text{GF}$ and the ghost $S_\text{GH}$ actions given by,

\begin{equation*}
    S_\text{GF} = \dfrac{1}{4\alpha} \ \chi^\alpha f_{\alpha\beta} \chi^\beta \hspace{1cm} \text{and,} \hspace{1cm}  S_\text{GH} = \Bar{c}^\alpha Q_{\alpha\beta} c^\beta,
\end{equation*} 

\noindent
respectively. Here, $\chi_\alpha = 0$ is the chosen gauge condition with $f_{\alpha\beta}$ any arbitrary symmetric, positive definite operator. The ghost action involves the operator, $Q^\alpha_\beta = \chi^\alpha_{,i}[\varphi] K^i_\beta[\varphi]$ and the ghost fields, $c$ and $\Bar{c}$ arising from the Faddeev-Popov method. Our main task is to evaluate \eqref{10} perturbatively in the orders of $\hbar$ up to the second order in $\hbar$. Here we present an expression for VD effective action up to two loops keeping the details of the derivation in Appendix \ref{AppA}. 

\noindent
The one-loop expression reads,

\begin{align}\label{11}
    \Gamma^{(1)}[\bar{\varphi}]=\lim_{\alpha\rightarrow 0}\left(\frac{i}{2}\text{ln det}S_{\text{GC} \ j}^{\ \ \ ;i}-i\text{ ln det} \bar Q^\alpha_\beta\right),
\end{align}

\noindent
where

\begin{align}\label{12}
    S_\text{GC} = \bar S[\varphi]+\frac{1}{4\alpha}\chi^\alpha f_{\alpha\beta} \chi^\beta.
\end{align}

\noindent
The expression for two-loop VD effective action reads (for details see Appendix \ref{AppA}),

\begin{align}\label{13}
        \Gamma^{(2)}[\Bar{\varphi}] &=  \Biggl \langle \dfrac{1}{4!} \Bar{S}_{;ijkl} \eta^i \eta^j \eta^k \eta^l + \dfrac{1}{4!} \Bar{S}_{\text{GF};ijkl} \eta^i \eta^j \eta^k \eta^l + \dfrac{1}{2} \Bar{S}_{;ij} \sigma^i_{(2)} \sigma^j_{(2)} + \dfrac{1}{2} \Bar{S}_{\text{GF};ij} \sigma^i_{(2)} \sigma^j_{(2)} + \nonumber \\& \hspace{7mm}  \dfrac{1}{2} \Bar{S}_{;ijk} \eta^i \eta^j \sigma^k_{(2)} + \dfrac{1}{2} \Bar{S}_{\text{GF};ijk} \eta^i \eta^j \sigma^k_{(2)} - \Bar{S}_{;ij} \eta^i \sigma^j_{(3)} - \Bar{S}_{\text{GF};ij} \eta^i \sigma^j_{(3)} +  \Bar{c}^\alpha \Bar{Q}_{\alpha\beta;i} \sigma^i_{(2)} c^\beta  \nonumber \\& \hspace{7mm} + \dfrac{1}{2} \Bar{c}^\alpha \Bar{Q}_{\alpha \beta;ij} \eta^i \eta^j c^\beta  - \Gamma^{(1)}_{;i} \sigma^i_{(2)} \Biggr \rangle  + \dfrac{i}{2} \Biggl \langle \Biggl( \dfrac{1}{3!} \Bar{S}_{;ijk} \eta^i \eta^j \eta^k +  \dfrac{1}{3!} \Bar{S}_{\text{GF};ijk} \eta^i \eta^j \eta^k + \nonumber \\& \hspace{7mm}\Bar{S}_{;ij} \eta^i \sigma^j_{(2)}  + \Bar{S}_{\text{GF};ij} \eta^i \sigma^j_{(2)} +  \Bar{c}^\alpha \Bar{Q}_{\alpha\beta;i} \eta^i c^\beta  - \Gamma^{(1)}_{;i} \eta^i \Biggr)^2 \Biggr \rangle.
    \end{align}

\noindent
The angular brackets in the expression above represent the expectation value in path integral formulation which can be computed by employing Wick’s theorem. The DeWitt choice has been made already, which is evident from the appearance of bars presiding over various quantities, which means evaluation in the limit $\varphi_*\rightarrow\bar\varphi$. It is essential that the derivatives be evaluated first and then DeWitt's choice be made. Also, the derivatives here are not simple functional derivatives. They are covariant functional derivatives calculated using the Vilkovisky-DeWitt connections $\Tilde{\Gamma}^i_{j k}$. The quantities $\sigma^i_{(n)}$ appear in the Taylor expansion of $\sigma^i[\varphi_*;\varphi]$ about $\varphi=\varphi_*$,

\begin{align}\label{14}
    \sigma^i[\varphi_*;\varphi]=-\eta^i+\sigma^i_{(2)}+\sigma^i_{(3)}+\mathcal{O}(\eta^4),
\end{align}

\noindent
where,   

\begin{align}\label{15}
    &\eta^i =  \varphi^i-\varphi_*^i\nonumber\\
    &\sigma^i_{(2)} = - \dfrac{1}{2} \Tilde{\Gamma}^i_{jk} \eta^j \eta^k \nonumber\\
    &\sigma^i_{(3)} = - \dfrac{1}{3!} \Bigl(\Tilde{\Gamma}^i_{j_1 j_2,j_3} + \Tilde{\Gamma}^i_{j_1 k} \Tilde{\Gamma}^k_{j_2 j_3} \Bigr) \eta^{j_1} \eta^{j_2} \eta^{j_3},
\end{align}

\section{Two-loop divergences in the Vilkovisky-DeWitt effective action}\label{SecIII}

\subsection{One-loop divergences}\label{SecIIIA}

\noindent
We observe that we require the VD connections $\Tilde{\Gamma}^i_{j k}$ to compute the derivatives appearing in the expression for effective action. In general, these connections are quite complicated to compute. But there exists a specific gauge, called the Landau-DeWitt gauge adopting, which allows us to replace the VD connections with the Christoffel ones. However, adopting the Landau-DeWitt gauge is beneficial only for the computation of one-loop VD effective action. We will see later that for two-loop computation, we would require full VD connections, although some simplification does take place on adopting the Landau-DeWitt gauge. Choosing the Landau-DeWitt gauge given by,

\begin{align}\label{16}
    \chi_\alpha = K^i_\alpha[\varphi_*]g_{ij}[\varphi_*](\varphi^j-\varphi^j_*)=0,
\end{align}

\noindent
the one loop VD effective action reads,

\begin{equation}\label{17}
     \Gamma^{(1)}[\Bar{\phi}] = - \text{ln det} \Bar{Q}_{\alpha\beta} + \lim_{\alpha \to 0} \ \dfrac{1}{2} \ \text{ln det}\left(\Bar{S}^{;i}_{j} + \dfrac{1}{2\alpha} \ \Bar{K}^{i}_{\beta} \Bar{K}^{\beta}_{j} \right).
 \end{equation}

 \noindent
In the above equation, all the covariant derivatives are to be calculated now using the configuration-space Christoffel connections $\Gamma^i_{jk}$.

\noindent
There are many places where the readers may find the explicit computation of one loop divergences in the VD effective action for theory \eqref{1}. In light of this, we only spent a few lines discussing the technique we will use here as well as for the two-loop computation later and writing down the one-loop divergences.

\noindent
Since our objective is to compute two loop divergences up to quadratic order in the background scalar field, we use the perturbative expansion technique. Note that the Schwinger-DeWitt technique can be used to get a closed-form expression for one loop divergences up to all orders in the background scalar field. However, since the technique becomes quite challenging to use when non-minimal operators are involved \cite{nonminimal}, we use a more approachable technique wherein we only keep terms up to second order in powers of background scalar field. Besides, we must not forget that the Schwinger-DeWitt technique is applicable exclusively for the computation of one-loop divergences only.

\noindent
The perturbative expansion technique can be applied for the computation of one-loop divergences as follows. We observe that the term in the brackets in \eqref{17} can be written as,

\begin{equation}\label{18}
     - \dfrac{1}{2} \ \text{ln det}\left(\Bar{S}^{;i}_{j} + \dfrac{1}{2\alpha} \ \Bar{K}^{i}_{\beta} \Bar{K}^{\beta}_{j} \right) = \ln{\int [d \eta] \exp \bigl\{ -S \bigr\}},
 \end{equation}

\noindent
where,

\begin{align}\label{19}
    &S = \dfrac{1}{2} S_{;ij} \eta^i \eta^j + \dfrac{1}{4 \alpha} K^\beta_i K_{j\beta} \eta^i \eta^j \nonumber \\
    &S_{;ij} = S_{,ij} - \tilde\Gamma^k_{ij} S_{,k}.
\end{align}

\noindent
We then expand S in \eqref{18} in the powers of background field as follows.

\begin{align}\label{20}
    \ln{\int [d \eta] \exp \bigl\{ -S \bigr\}} &= \ln{\int [d \eta] \exp \bigl\{ -S_0 - S_1 - S_2\bigr\}} \nonumber \\
    &= \ln{\int [d \eta] \exp \bigl\{ -S_0\bigr\}} \Big( 1 - S_1 - S_2 + \dfrac{(S_1)^2}{2}\Big) \nonumber \\
    &= \ln\Big( 1 - \bigl\langle S_1 \bigr\rangle - \bigl\langle S_2 \bigr\rangle + \dfrac{\bigl\langle (S_1)^2 \bigr\rangle}{2}\Big) \nonumber \\ 
    &= - \bigl\langle S_1 \bigr\rangle - \bigl\langle S_2 \bigr\rangle + \dfrac{\bigl\langle (S_1)^2 \bigr\rangle}{2}.
\end{align}

\noindent
The subscript `$n$' in $S_{n}$ stands for the order of the background scalar field. For example, the term $S_0$ is zeroth order in the background field, and it gives us the propagators. The angular brackets represent the expectation value in path integral formulation which can be computed by employing Wick’s theorem. On using \eqref{20} in \eqref{17} we obtain,
    
\begin{equation}\label{21}
     \Gamma^{(1)}[\Bar{\phi}] = - \text{ln det} \Bar{Q}_{\alpha\beta} + \lim_{\alpha \to 0} \ \biggl( \bigl\langle S_1 \bigr\rangle + \bigl\langle S_2 \bigr\rangle - \dfrac{\bigl\langle (S_1)^2 \bigr\rangle}{2} \biggr).
 \end{equation}

\noindent  
The computation of one-loop divergences in VD effective action following the technique outlined above runs exactly in line with \cite{mackaytoms, aashish2021covariant,sandeep,ourpaper2}. For the background comprising of a constant scalar field and flat Minkowski metric, we have the following result for one-loop divergences in VD effective action,

\begin{equation}\label{22}
    \Gamma^{(1)}_{div} = \dfrac{5 m^4 \kappa^2}{128 \pi^2 \epsilon} \int d^4x \sqrt{g(x)} \ \phi^2.
\end{equation}

\subsection{Vilkovisky-DeWitt Connections}\label{SecIIIB}

\noindent
The expression for two-loop VD effective action \eqref{13} involves the VD connections at various places. These connections arise in the expression for $\sigma^i_{(n)}$ appearing in the Taylor expansion of $\sigma^i[\varphi_*;\varphi]$ as well as in the covariant derivatives of action. One cannot simply replace the VD connections with the Christoffel ones everywhere. However, it may be possible for us to use this replacement in some places if we adopt the Landau-DeWitt gauge. To see this, we first write down the expression for VD connection,

\begin{equation}\label{23}
    \Tilde{\Gamma}^i_{jk} = \Gamma^i_{jk} + \dfrac{1}{2} \gamma^{\alpha\epsilon} \gamma^{\beta\sigma} K_{\alpha j} K_{\beta k} \Bigl( K^n_\epsilon K^i_{\sigma;n} + K^n_\sigma K^i_{\epsilon;n} \Bigr) - \gamma^{\alpha \beta} \Bigl( K_{\alpha j} K^i_{\beta;k} + K_{\alpha k} K^i_{\beta;j} \Bigr),
\end{equation}

\noindent
where $\Gamma^i_{jk}$ are the Christoffel connections. 

\noindent
Let us see how the expression for $\sigma^i_{(2)}$ is simplified with the Landau-DeWitt gauge choice. The expression for $\sigma^i_{(2)}$ \eqref{15} involves two $\eta$'s
contracted with the VD connection. Since the VD connections are proportional to $K_{\alpha i}$ with free Latin index, the expression for $\sigma^i_{(2)}$ will involve products such as $K_{\alpha i}\eta^i$ which vanish if we choose the Landau-DeWitt gauge. Thus the extra terms in \eqref{23} besides the Christoffel connection can be safely dropped while evaluating $\sigma^i_{(2)}$. The simplification carried out above is to be used wherever applicable in the expression for two-loop VD effective action. 

\noindent
The expression for $\sigma^i_{(3)}$ \eqref{15} involves a product of two VD connections, and only the one which is contracted with two $\eta$'s can be replaced with Christoffel connection. The products of VD connections also appear in various covariant derivatives of $S$, and the same simplification follows there too. Additionally, we also have a once differentiated VD connection term in the expression for $\sigma^i_{(3)}$ and even twice differentiated VD connection in the higher order covariant derivatives of $S$.
Depending upon how these differentiated VD connections are contracted with $\eta$'s, these, too, can be replaced with the Christoffel connections. The simplifications that follow with the use of the Landau-DeWitt gauge have been explained thoroughly in Appendix \ref{AppB}

\noindent
 We remind ourselves that the small Latin letters attached to quantities stand for a group of indices as well as the space-time argument. Since we are interested in the computation of two-loop divergences, in theory, \eqref{1}, which involves two fields, namely, a scalar field and the metric field, the small Latin indices may either represent space-time indices for a tensor field or for a scalar field. For example, we have, in general, six possible connection terms like: ($\Gamma^{g_{\mu\nu}(x)}_{\phi(x')g_{\alpha\beta}(x'')}$, $\Gamma^{\phi(x)}_{g_{\mu\nu}(x')g_{\alpha\beta}(x'')}$ and so on). However, a neater way would be to use the following notation: $g_{\mu\nu}\rightarrow 1$, $\phi\rightarrow 2$. For example, in this notation we would write $\Gamma^{g_{\mu\nu}(x)}_{\phi(x')g_{\alpha\beta}(x'')}$ succinctly as $(\Gamma^1_{12})^{\alpha\beta}_{\mu\nu}(x,x',x'')$. The numerals are only placeholders for the fields. The actual space-time indices associated with the fields are to be attached appropriately with the quantity having small Latin indices, as we have shown above. We will be following this notation only when we wish to write explicitly each component of a quantity carrying small Latin letters. To this end, we have written down the expression for VD connections which may be found in earlier papers \cite{cho1989vilkovisky,cho1991vilkovisky,odintsov1993gaugeh},

\begin{align}\label{24}
    &\bigl(\Tilde{\Gamma}^1_{11}\bigr)^{\mu \nu \rho \sigma}_{\lambda \tau}(x,x',x'') = \delta(x,x') \delta(x,x'') \Biggl( -\delta^{(\mu}_{(\lambda} \eta^{\nu)(\rho}\delta^{\sigma)}_{\tau)} + \dfrac{1}{4} \eta^{\mu \nu}\delta^\rho_{(\lambda}\delta^\sigma_{\tau)}  + \dfrac{1}{4} \eta^{\rho \sigma}\delta^\mu_{(\lambda}\delta^\nu_{\tau)}  + \nonumber \\& \hspace{4.5cm}\dfrac{1}{4} \eta_{\lambda \tau} \eta^{\mu (\rho}\eta^{\sigma)\nu} -  \dfrac{1}{8} \eta_{\lambda \tau} \eta^{\mu \nu} \eta^{\rho \sigma} \Biggr)   - \dfrac{\delta(x,x')}{4} \Biggl(\dfrac{\eta^{\mu (\rho} \partial^{\sigma)} \partial^\nu}{\Box} +  \nonumber \\& \hspace{4.5cm}  \dfrac{\eta^{\nu (\rho} \partial^{\sigma)} \partial^\mu}{\Box} - \eta^{\mu \nu} \dfrac{\partial^\rho \partial^\sigma}{\Box} -  \eta^{\rho \sigma} \dfrac{\partial^\mu \partial^\nu}{\Box} + \dfrac{\eta^{\mu \nu} \eta^{\rho \sigma}}{2} \Biggr) \delta(x,x'') \nonumber \\
    &\bigl(\Tilde{\Gamma}^1_{22}\bigr)_{\mu \nu}(x,x',x'') = \dfrac{\eta_{\mu \nu}}{4} \ \delta(x,x') \delta(x,x'') \nonumber \\
    &\bigl(\Tilde{\Gamma}^2_{12}\bigr)^{\mu \nu}(x,x',x'') = \dfrac{\delta(x,x'')}{2} \dfrac{\partial^\mu \partial^\nu}{\Box} \delta(x,x'),
\end{align}

\noindent
where $\eta_{\mu\nu}$ is the Minkowski metric in accordance with our choice of the background fields. (Note that the connections are to be computed for background fields in \eqref{24})

\subsection{Two-loop divergences}\label{SecIIIC}

\noindent
We wish to carry out an explicit computation of \eqref{13} for the theory in \eqref{1} in this section. The covariant derivatives of $S$ appearing in \eqref{13} are to be expanded in a usual way using the VD connections at some places and Christoffel ones elsewhere following the simplification carried out in Appendix \ref{AppB} in the Landau-DeWitt gauge. For the theory of our interest $\eta^i = \{h_{\mu \nu}, \psi \}$ is the quantum fluctuation around the background, where $h_{\mu\nu}$ is the fluctuation around the Minkowski metric and $\psi$ is the fluctuation around a constant massive scalar field. To explicitly expand out the summation over dummy indices, we make use of the notation introduced at the end of \ref{SecIIIB}. The ordinary functional derivatives that are left behind after the expansion are to be evaluated at $\varphi^i = \bar\varphi^i$. The double expansion to be carried out, first the expansion of covariant derivatives and then the expansion of summation over dummy indices, is quite cumbersome and lengthy but straightforward. In light of this, we skip carrying out the expansion explicitly here but show the averages of the quantum fields to be evaluated and the loop integrals that follow from these in the momentum space. 

\noindent
We begin by writing down the propagators for the scalar field and graviton in flat Minkowski background,

\begin{align}\label{25}
    &\Bigl \langle\psi(x) \psi(x')\Bigr \rangle = G(x,x') = \int \dfrac{d^4k}{(2 \pi)^4} \ \Tilde{G}(k) \  e^{i k (x-x')} \nonumber \\
    &\Bigl \langle h_{\mu \nu}(x) h_{\rho \sigma}(x')\Bigr \rangle = G_{\mu \nu \rho \sigma}(x,x') = \int \dfrac{d^4k}{(2 \pi)^4} \ \Tilde{G}_{\mu \nu \rho \sigma}(k) \ e^{i k (x-x')},
\end{align}

\noindent
with

\begin{align}\label{26}
    &\Tilde{G}(k) = \dfrac{i}{k^2 + m^2} \nonumber \\
    &\Tilde{G}_{\mu \nu \rho \sigma}(k) = \dfrac{i}{2 k^2} \bigl(g_{\mu \rho} g_{\nu \sigma} + g_{\mu \sigma} g_{\nu \rho} - g_{\mu \nu} g_{\rho \sigma}\bigr) + \dfrac{i (\alpha - 1)}{2 k^4} \bigl( g_{\mu \rho} k_\nu k_\sigma + g_{\mu \sigma} k_\nu k_\rho + g_{\nu \rho} k_\mu k_\sigma \nonumber \\& \hspace{2cm} + g_{\nu \sigma} k_\mu k_\rho \bigr).
\end{align}

\noindent
Additionally, we also have the ghost field propagator given by,

\begin{align}\label{27}
    \Bigl \langle \Bar{c}_\mu(x) c_\nu(x')\Bigr \rangle = G^\text{Ghost}_{\mu \nu}(x,x') = \int \dfrac{d^4k}{(2 \pi)^4} \ \Tilde{G}^\text{Ghost}_{\mu \nu}(k) \  e^{i k (x-x')},
\end{align}

\noindent
with

\begin{align}\label{28}
    \Tilde{G}^\text{Ghost}_{\mu \nu}(k) = \dfrac{i g_{\mu \nu}}{k^2}.
\end{align}

\noindent
In the explicit calculation of \eqref{13} we encounter in general, $n$ point functions $\langle\eta^{i_1}\eta^{i_2}...\eta^{i_n}\rangle$. These can be broken down to a sum of products of 2-point functions using Wick's theorem, after which we employ \eqref{25} and \eqref{27}.

\subsubsection{Loop integrals and divergent parts}

\begin{figure}[h!]
    \centering
    \subfloat[\centering ]
    {\begin{tikzpicture}
    \begin{feynman}
    \node [dot,label=left:\(x\)] (a) at (0,0);
    \vertex (b) at (2,0); \vertex (c) at (-2,0);
    \diagram*{
    (a) -- [red, gluon, half left] (b) -- [red, gluon, half left] (a) -- [red, gluon, half left] (c) -- [red, gluon, half left] (a)
    };
    \end{feynman}
    \end{tikzpicture}} %
    \qquad
    \subfloat[\centering ]
    {\begin{tikzpicture}
    \begin{feynman}
    \node [dot,label=left:\(x\)] (a) at (0,0);
    \vertex (b) at (2,0); \vertex (c) at (-2,0);
    \diagram*{
    (a) -- [red, gluon, half left] (b) -- [red, gluon, half left] (a) -- [blue, half left] (c) -- [blue, half left] (a)
    };
    \end{feynman}
    \end{tikzpicture}}
    \qquad
    \subfloat[\centering ]
    {\begin{tikzpicture}
    \begin{feynman}
    \node [dot,label=left:\(x\)] (a) at (0,0);
    \vertex (b) at (2,0); \vertex (c) at (-2,0);
    \diagram*{
    (a) -- [blue, half left] (b) -- [blue, half left] (a) -- [blue, half left] (c) -- [blue, half left] (a)
    };
    \end{feynman}
    \end{tikzpicture}}
    \qquad
    \subfloat[\centering ]
    {\begin{tikzpicture}
    \begin{feynman}
    \node [dot,label=left:\(x\)] (a) at (0,0);
    \vertex (b) at (2,0); \vertex (c) at (-2,0);
    \diagram*{
    (a) -- [scalar, half left] (b) -- [scalar, half left] (a) -- [red, gluon, half left] (c) -- [red, gluon, half left] (a)
    };
    \end{feynman}
    \end{tikzpicture}}
    \qquad
    \subfloat[\centering ]
    {\begin{tikzpicture}
    \begin{feynman}
    \node [dot,label=left:\(x\)] (a) at (0,0);
    \vertex (b) at (2,0); \vertex (c) at (-2,0);
    \diagram*{
    (a) -- [scalar, half left] (b) -- [scalar, half left] (a) -- [blue, half left] (c) -- [blue, half left] (a)
    };
    \end{feynman}
    \end{tikzpicture}}
    
    \caption{\label{fig:4.1}  Loop diagrams for $I_1$}
\end{figure}
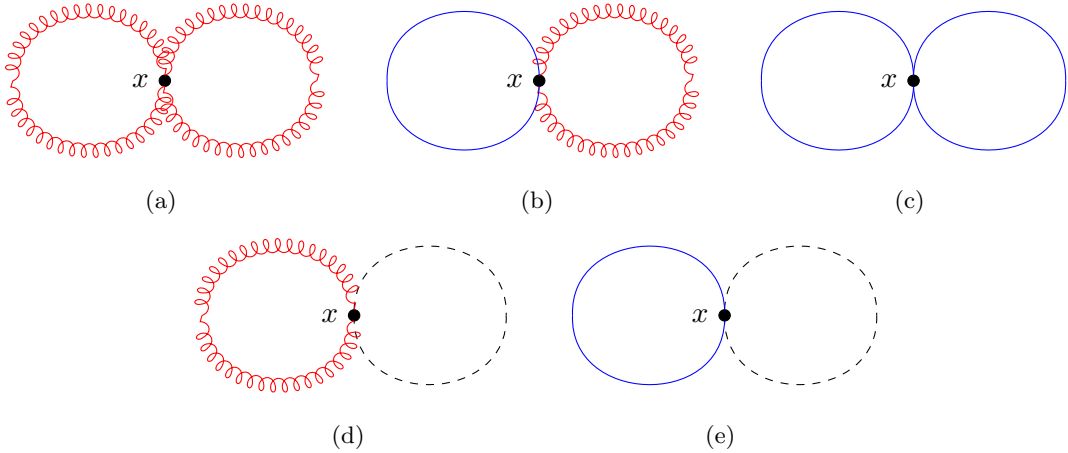

\noindent
The expression \eqref{13} reads,

\begin{align}\label{29}
        \Gamma^{(2)}[\Bar{\varphi}] &=  \Biggl \langle \dfrac{1}{4!} \Bar{S}_{;ijkl} \eta^i \eta^j \eta^k \eta^l + \dfrac{1}{4!} \Bar{S}_{\text{GF};ijkl} \eta^i \eta^j \eta^k \eta^l + \dfrac{1}{2} \Bar{S}_{;ij} \sigma^i_{(2)} \sigma^j_{(2)} + \dfrac{1}{2} \Bar{S}_{\text{GF};ij} \sigma^i_{(2)} \sigma^j_{(2)} + \nonumber \\& \hspace{7mm} \dfrac{1}{2} \Bar{S}_{;ijk} \eta^i \eta^j \sigma^k_{(2)} + \dfrac{1}{2} \Bar{S}_{\text{GF};ijk} \eta^i \eta^j \sigma^k_{(2)} - \Bar{S}_{;ij} \eta^i \sigma^j_{(3)} - \Bar{S}_{\text{GF};ij} \eta^i \sigma^j_{(3)} +  \Bar{c}^\alpha \Bar{Q}_{\alpha\beta;i} \sigma^i_{(2)} c^\beta  \nonumber \\& \hspace{7mm} + \dfrac{1}{2} \Bar{c}^\alpha \Bar{Q}_{\alpha \beta;ij} \eta^i \eta^j c^\beta  - \Gamma^{(1)}_{;i} \sigma^i_{(2)} \Biggr \rangle  + \dfrac{i}{2} \Biggl \langle \Biggl( \dfrac{1}{3!} \Bar{S}_{;ijk} \eta^i \eta^j \eta^k +  \dfrac{1}{3!} \Bar{S}_{\text{GF};ijk} \eta^i \eta^j \eta^k +  \nonumber \\& \hspace{7mm} \Bar{S}_{;ij} \eta^i \sigma^j_{(2)} + \Bar{S}_{\text{GF};ij} \eta^i \sigma^j_{(2)} +  \Bar{c}^\alpha \Bar{Q}_{\alpha\beta;i} \eta^i c^\beta  - \Gamma^{(1)}_{;i} \eta^i \Biggr)^2 \Biggr \rangle.
    \end{align}

\noindent
Let's begin with,

\begin{align}\label{30}
    I_1 &= \Biggl \langle \dfrac{1}{4!} \Bar{S}_{;ijkl} \eta^i \eta^j \eta^k \eta^l + \dfrac{1}{4!} \Bar{S}_{\text{GF};ijkl} \eta^i \eta^j \eta^k \eta^l + \dfrac{1}{2} \Bar{S}_{;ij} \sigma^i_{(2)} \sigma^j_{(2)} + \dfrac{1}{2} \Bar{S}_{\text{GF};ij} \sigma^i_{(2)} \sigma^j_{(2)} + \nonumber \\& \dfrac{1}{2} \Bar{S}_{;ijk} \eta^i \eta^j \sigma^k_{(2)} +   \dfrac{1}{2} \Bar{S}_{\text{GF};ijk} \eta^i \eta^j \sigma^k_{(2)} - \Bar{S}_{;ij} \eta^i \sigma^j_{(3)} - \Bar{S}_{\text{GF};ij} \eta^i \sigma^j_{(3)} +  \Bar{c}^\alpha \Bar{Q}_{\alpha\beta;i} \sigma^i_{(2)} c^\beta + \nonumber \\& \dfrac{1}{2} \Bar{c}^\alpha \Bar{Q}_{\alpha \beta;ij} \eta^i \eta^j c^\beta  - \Gamma^{(1)}_{;i} \sigma^i_{(2)} \Biggr \rangle.
\end{align}

\noindent
The expansion has been carried out in MATHEMATICA. We encounter various four point functions, namely:: $\Bigl \langle h(x) h(x) h(x) h(x) \Bigr \rangle$, $\Bigl \langle h(x) h(x) \psi(x) \psi(x) \Bigr \rangle$, $\Bigl \langle \psi(x) \psi(x) \psi(x) \psi(x) \Bigr \rangle$, \\ $\Bigl \langle h(x) h(x) h(x) \psi(x) \Bigr \rangle$, $\Bigl \langle \psi(x) \psi(x) \psi(x) h(x) \Bigr \rangle$, $\Bigl \langle h(x) h(x) \Bar{c}^\alpha(x) c_\beta(x) \Bigr \rangle$, and \\ $\Bigl \langle \psi(x) \psi(x) \Bar{c}^\alpha(x) c_\beta(x) \Bigr \rangle$, where we have suppressed the space-time indices of $h_{\mu\nu}$ for brevity. From Wick's theorem, it is evident that odd-point functions vanish; thus $\Bigl \langle h h h \psi \Bigr \rangle$, and $\Bigl \langle \psi \psi \psi h \Bigr \rangle$ four-point functions vanish. The remaining ones correspond to the loop diagrams (a), (b), (c), (d), and (e) in Fig:\ref{fig:4.1}, where the blue line represents the scalar field propagator, the red spiral represents the graviton propagator, and the black dashed line represents the ghost field propagator.

\noindent
Note that in \eqref{30}, we have a term involving one-loop effective action,

\begin{equation}\label{31}
     \Bar{\Gamma}^{(1)}_{,i} \sigma^i_{(2)} = - \dfrac{1}{2} \Bar{\Gamma}^{(1)}_{,i} \Gamma^i_{jk} \eta^j \eta^k,
\end{equation}

\noindent
where $\Bar{\Gamma}^{(1)}$ is one-loop effective action. The divergences in one-loop effective action for the theory of scalar field non-minimally coupled to gravity  have been computed in \cite{genback} for arbitrary metric and scalar field background. Choosing an arbitrary background for the metric and scalar field is an essential requirement for one loop effective action since we need to compute the derivatives prior to setting $\varphi^i=\bar\varphi^i$. Setting the non-minimal coupling parameter to zero and turning off the quartic scalar self-coupling in the divergent part of one loop effective action, we have

\begin{align}\label{32}
    \Bar{\Gamma}^{(1)} &= \sqrt{g} \ \Biggl( \dfrac{5 m^4 \phi^2}{128 \pi^2 \epsilon} + \dfrac{13 m^2 \phi^2 R}{192 \pi^2 \epsilon} + \dfrac{5 m^2 \phi \nabla_\mu \nabla^\mu \phi}{128 \pi^2 \epsilon} + \dfrac{\phi R \nabla_\mu \nabla^\mu \phi}{192 \pi^2 \epsilon} - \dfrac{R^{\mu \nu} \phi \nabla_\mu \nabla_\nu\phi}{8 \pi^2 \epsilon} - \nonumber \\& \dfrac{9 \phi \nabla^\mu \nabla_\mu \nabla^\nu \nabla_\nu \phi}{128 \pi^2 \epsilon} \Biggr).
\end{align}

\noindent
The derivatives with respect to the metric and scalar field are respectively given by,

\begin{align}\label{33}
    &(\Bar{\Gamma}^{(1)}_{,1})^{\mu\nu} = \dfrac{5 m^4 \phi^2 \eta^{\mu\nu}}{256 \pi^2 \epsilon} \nonumber \\
    &\Bar{\Gamma}^{(1)}_{,2} = \dfrac{5 m^4 \phi}{64 \pi^2 \epsilon},
\end{align}

\noindent
where the derivatives have been computed at the chosen background fields. 

\noindent
Let us focus on terms involving covariant derivatives of $S$. These can be broadly categorized into terms involving only the Christoffel connections and those involving VD connections. 

\begin{itemize}

    \item The terms involving Christoffel connections are straightforward to deal with. For example, the term \\ $\Bar{S}_{,mij} \Gamma^m_{kl}  \eta^i \eta^j \eta^k \eta^l $ when expanded out by summing over dummy indices involves a sum of terms of following type,
    
\begin{align}\label{34}
    \int d^4x f(x) A(x)B(x)C(x)D(x).
\end{align}

The quantities $A(x), B(x), C(x)$ and $D(x)$ are just placeholders for either of the quantum fields $\{h_{\mu\nu},\psi\}$. Note that according to the DeWitt notation, each Latin index comprises space-time indices associated with the field as well the space-time argument of the field. Each distinct Latin index corresponds to a distinct set of space-time indices and space-time arguments (for example, if $i$ and $j$ correspond to a tensor field, then: $i\equiv\{\mu\nu, x\}$ and $j\equiv\{\alpha\beta,x'\}$). We have also assumed the following while writing down \eqref{34},

\begin{align}\label{35}
    \bar S_{,ijk} = f(x)\delta(x,x')\delta(x,x''),
\end{align}

where we have suppressed space-time indices associated with the fields. For example if $i\equiv\{\mu\nu,x\}, j\equiv\{\alpha\beta,x'\}$ and $k\equiv\{x''\}$ then $\bar S_{,ijk}$ would read,

\begin{align}\label{36}
    \bar S_{,ijk} = f^{\mu\nu\alpha\beta}(x)\delta(x,x')\delta(x,x'').
\end{align}

The quantum average of \eqref{34} can then be decomposed using Wick's theorem as follows,

\begin{align}\label{37}
    \langle A(x)B(x)C(x)D(x)\rangle &= \langle A(x)B(x)\rangle\langle C(x)D(x)\rangle + \langle A(x)C(x)\rangle\langle B(x)D(x)\rangle + \nonumber \\& \hspace{7mm} \langle A(x)D(x)\rangle\langle B(x)C(x)\rangle.
\end{align}

Each term in the sum above can be Fourier transformed, followed by the use of propagators. For example for scalar fields($A(x)=\psi(x)$, $B(x)=\psi(x)$),

\begin{align}\label{38}
    \langle A(x)B(x)\rangle = \langle \psi(x)\psi(x)\rangle = \int \frac{d^4k}{(2\pi)^4}\tilde G(k).
\end{align}

\item The terms involving VD connections require further work owing to the presence of $\frac{\partial\mu\partial_\nu}{\Box}\delta(x,x')$ in the expression of VD connections. These can further be categorized based on which VD connection is involved (Note that there are only two VD connections, namely, $\Tilde\Gamma^1_{11}$ and $\Tilde\Gamma^2_{12}$ that involve $\frac{1}{\Box}$). Let us examine each of these cases. In what follows, we will focus entirely on the additional piece that involves $\frac{\partial\mu\partial_\nu}{\Box}\delta(x,x')$ in the term $\Bar{S}_{,ni} \Gamma^m_{jk} \Tilde{\Gamma}^n_{lm} \eta^i \eta^j \eta^k \eta^l$.

\begin{itemize}
    \item[\Large{$\ast$}] If the connection involved is $\Tilde\Gamma^1_{11}$, then we have, 

\begin{align}\label{39}
    &\int d^4x d^4x' d^4x_1 d^4x_2 d^4x_3 d^4x_4 f(x) \delta(x,x_1) \delta(x,x_2) \Bigl(\dfrac{\partial'^\mu \partial'^\nu}{\Box'} \delta(x_2,x') \Bigr) \times \nonumber \\& \delta(x_3,x') \delta(x_4,x') 
    A(x_1) B(x_2) C(x_3) D(x_4) \nonumber \\
    &\hspace{-5mm}= \int d^4x d^4x' f(x)\Bigl(\dfrac{\partial'^\mu \partial'^\nu}{\Box'} \delta(x,x') \Bigr) A(x) B(x') C(x')D(x).
\end{align}

On using Wick's theorem, we find,

\begin{align}\label{40}
    \int d^4x d^4x' f(x)\Bigl(\dfrac{\partial'^\mu \partial'^\nu}{\Box'} \delta(x,x') \Bigr) \Big(\langle A(x) B(x')\rangle \langle C(x')D(x)\rangle + \nonumber\\& \hspace{-9cm} \langle A(x) C(x')\rangle\langle B(x')D(x)\rangle  + \langle A(x) D(x)\rangle\langle B(x')C(x')\rangle\Big).
\end{align}

We will show that the last term in the expression above involving \\ $\langle A(x) D(x)\rangle\langle B(x')C(x')\rangle$ will not make any contribution if a constant scalar field is chosen as the background.
The first two terms in \eqref{40} can be dealt with in a straightforward manner by making use of the Fourier transform of the Dirac delta function,

\begin{equation}\label{41}
    \delta(x,x') = \int \dfrac{d^4 k}{(2 \pi)^4} e^{i k (x - x')}.
\end{equation}

The first term in \eqref{40} reads,

\begin{align}\label{42}
    &\int d^4x d^4x' f(x)\Bigl(\dfrac{\partial'^\mu \partial'^\nu}{\Box'} \delta(x,x') \Bigr) \langle A(x) B(x')\rangle\langle C(x')D(x)\rangle \nonumber\\&= \int \frac{d^4k_1}{(2\pi)^4}\int \frac{d^4k_2}{(2\pi)^4}\frac{(k^\mu_1+k^\mu_2)(k^\nu_1+k^\nu_2)}{(k_1+k_2)^2}\tilde G_A(k_1)\tilde G_C(k_2),
\end{align}

where $\tilde G_A(k)$ and $\tilde G_C(k)$ are the momentum space propagators for the fields $A$ and $C$ respectively. The second term in \eqref{40} can also be dealt with in a similar manner. The third term in \eqref{40} reads,

\begin{align}\label{43}
     &\int d^4x d^4x' f(x)\Bigl(\dfrac{\partial'^\mu \partial'^\nu}{\Box'} \delta(x,x') \Bigr) \langle A(x) D(x)\rangle\langle B(x')C(x')\rangle \nonumber\\&= \int d^4x f(x)\int d^4x' \Bigl(\dfrac{\partial'^\mu \partial'^\nu}{\Box'} \delta(x,x') \Bigr) \int \frac{d^4k_1}{(2\pi)^4}G_A(k_1)\int \frac{d^4k_2}{(2\pi)^4}G_C(k_2)\nonumber\\
     &=\int d^4x\Bigl(\dfrac{\partial^\mu \partial^\nu}{\Box} f(x) \Bigr)\int \frac{d^4k_1}{(2\pi)^4}G_A(k_1)\int \frac{d^4k_2}{(2\pi)^4}G_C(k_2).
\end{align}

The function $f(x)$ may either be a constant or involve some function of the background scalar field in general. If the chosen background is a constant scalar field, then it is clear that \eqref{43} vanishes.

\item[\Large{$\ast$}] If the connection involved is $\Tilde\Gamma^2_{12}$, we have,

\begin{align}\label{44}
     \int d^4x f(x)\int d^4 x' \Bigl \langle \Bigl(\dfrac{\partial^\mu \partial^\nu}{\Box} \delta(x,x') \Bigr) A(x)& B(x) C(x) D(x') \Bigr \rangle = \int d^4x f(x) \times \nonumber \\& \hspace{-5cm} \int \dfrac{d^4 k_1}{(2 \pi)^4} \dfrac{d^4 k_2}{(2 \pi)^4}  \Biggl[G_A(k_1)\dfrac{k^\mu_2 k^\nu_2}{k^2_2}G_C(k_2)  + G_A(k_1)\dfrac{k^\mu_2 k^\nu_2}{k^2_2}G_B(k_2)  +  \nonumber \\& \hspace{-5cm} G_B(k_1)\dfrac{k^\mu_2 k^\nu_2}{k^2_2} G_A(k_2)  \Biggr],
\end{align}

\noindent
where all the terms contribute independent of $f(x)$, unlike the previous case.

\end{itemize} 

\end{itemize}

\noindent
Once the expression for $I_1$ \eqref{30} is broken down following the method outlined above, it is followed by the substitution of expressions for propagators in momentum space, leaving to us a set of divergent integrals listed below,

\begin{align}\label{45}
    &(1) \int \dfrac{d^4 p_1}{(2 \pi)^4} \dfrac{d^4 p_2}{(2 \pi)^4} \ \dfrac{1}{p^2_1 p^2_2} \nonumber \\
    &(2) \int \dfrac{d^4 p_1}{(2 \pi)^4} \dfrac{d^4 p_2}{(2 \pi)^4} \ \dfrac{p^a_1 p^b_1 p_{2a} p_{2b}}{p^4_1 p^4_2} \nonumber \\
    &(3) \int \dfrac{d^4 p_1}{(2 \pi)^4} \dfrac{d^4 p_2}{(2 \pi)^4} \ \dfrac{1}{p^2_1 (p^2_2 + m^2)} \hspace{1cm}  \int \dfrac{d^4 p_1}{(2 \pi)^4} \dfrac{d^4 p_2}{(2 \pi)^4} \ \dfrac{1}{(p^2_1 + m^2) p^2_2} \nonumber \\
    &(4) \int \dfrac{d^4 p_1}{(2 \pi)^4} \dfrac{d^4 p_2}{(2 \pi)^4} \ \dfrac{1}{(p^2_1 + m^2) (p^2_2 + m^2)} \nonumber \\
    &(5) \int \dfrac{d^4 p_1}{(2 \pi)^4} \dfrac{d^4 p_2}{(2 \pi)^4} \ \dfrac{1}{p^2_1 (p_1 + p_2)^2} \hspace{1cm}  \int \dfrac{d^4 p_1}{(2 \pi)^4} \dfrac{d^4 p_2}{(2 \pi)^4} \ \dfrac{1}{p^2_2 (p_1 + p_2)^2} \nonumber \\
    &(6) \int \dfrac{d^4 p_1}{(2 \pi)^4} \dfrac{d^4 p_2}{(2 \pi)^4} \ \dfrac{p^a_1 p_{2a}}{p^2_1 p^2_2 (p_1 + p_2)^2} \nonumber \\
    &(7) \int \dfrac{d^4 p_1}{(2 \pi)^4} \dfrac{d^4 p_2}{(2 \pi)^4} \ \dfrac{p^a_1 p^b_1 p_{2a} p_{2b}}{p^2_1 p^4_2 (p_1 + p_2)^2} \hspace{1cm}  \int \dfrac{d^4 p_1}{(2 \pi)^4} \dfrac{d^4 p_2}{(2 \pi)^4} \ \dfrac{p^a_1 p^b_1 p_{2a} p_{2b}}{p^4_1 p^2_2 (p_1 + p_2)^2} \nonumber \\
    &(8) \int \dfrac{d^4 p_1}{(2 \pi)^4} \dfrac{d^4 p_2}{(2 \pi)^4} \ \dfrac{p^a_1 p^b_1 p_{2a} p_{2b}}{p^6_1 p^2_2 (p_1 + p_2)^2} \hspace{1cm}  \int \dfrac{d^4 p_1}{(2 \pi)^4} \dfrac{d^4 p_2}{(2 \pi)^4} \ \dfrac{p^a_1 p^b_1 p_{2a} p_{2b}}{p^2_1 p^6_2 (p_1 + p_2)^2} \nonumber \\
    &(9) \int \dfrac{d^4 p_1}{(2 \pi)^4} \dfrac{d^4 p_2}{(2 \pi)^4} \ \dfrac{p^a_1 p_{2a}}{p^4_1 p^2_2 (p_1 + p_2)^2} \hspace{1cm} \int \dfrac{d^4 p_1}{(2 \pi)^4} \dfrac{d^4 p_2}{(2 \pi)^4} \ \dfrac{p^a_1 p_{2a}}{p^2_1 p^4_2 (p_1 + p_2)^2} \nonumber \\
    &(10) \int \dfrac{d^4 p_1}{(2 \pi)^4} \dfrac{d^4 p_2}{(2 \pi)^4} \ \dfrac{1}{p^2_1 p^2_2 (p_1 + p_2)^2} \nonumber \\
    &(11) \int \dfrac{d^4 p_1}{(2 \pi)^4} \dfrac{d^4 p_2}{(2 \pi)^4} \ \dfrac{p^a_1 p^b_1 p_{2a} p_{2b}}{p^4_1 (p^2_2 + m^2) (p_1 + p_2)^2} \hspace{1cm} \int \dfrac{d^4 p_1}{(2 \pi)^4} \dfrac{d^4 p_2}{(2 \pi)^4} \ \dfrac{p^a_1 p^b_1 p_{2a} p_{2b}}{p^4_2 (p^2_1 + m^2) (p_1 + p_2)^2} \nonumber \\
    &(12) \int \dfrac{d^4 p_1}{(2 \pi)^4} \dfrac{d^4 p_2}{(2 \pi)^4} \ \dfrac{p^a_1 p_{2a} }{p^2_1 (p^2_2 + m^2) (p_1 + p_2)^2} \hspace{1cm} \int \dfrac{d^4 p_1}{(2 \pi)^4} \dfrac{d^4 p_2}{(2 \pi)^4} \ \dfrac{p^a_1 p_{2a}}{p^2_2 (p^2_1 + m^2) (p_1 + p_2)^2} \nonumber \\
    &(13) \int \dfrac{d^4 p_1}{(2 \pi)^4} \dfrac{d^4 p_2}{(2 \pi)^4} \ \dfrac{p^a_1 p^b_1 p^c_1 p_{2a} p_{2b} p_{2c}}{p^4_1 p^4_2 (p_1 + p_2)^2} \nonumber \\
    &(14) \int \dfrac{d^4 p_1}{(2 \pi)^4} \dfrac{d^4 p_2}{(2 \pi)^4} \ \dfrac{p^2_1}{(p^2_1 + m^2) (p^2_2 + m^2)} \hspace{1cm} \int \dfrac{d^4 p_1}{(2 \pi)^4} \dfrac{d^4 p_2}{(2 \pi)^4} \ \dfrac{p^2_2}{(p^2_1 + m^2) (p^2_2 + m^2)} \nonumber \\
    &(15) \int \dfrac{d^4 p_1}{(2 \pi)^4} \dfrac{d^4 p_2}{(2 \pi)^4} \ \dfrac{p^a_1 p_{2a}}{(p^2_1 + m^2) (p^2_2 + m^2)} \nonumber \\
    &(16) \int \dfrac{d^4 p_1}{(2 \pi)^4} \dfrac{d^4 p_2}{(2 \pi)^4} \ \dfrac{p^2_1}{p^2_2 (p_1 + p_2)^2} \hspace{1cm} \int \dfrac{d^4 p_1}{(2 \pi)^4} \dfrac{d^4 p_2}{(2 \pi)^4} \ \dfrac{p^2_2}{p^2_1 (p_1 + p_2)^2} \nonumber \\
    &(17) \int \dfrac{d^4 p_1}{(2 \pi)^4} \dfrac{d^4 p_2}{(2 \pi)^4} \ \dfrac{p^a_1 p_{2a}}{p^2_1 (p_1 + p_2)^2} \hspace{1cm} \int \dfrac{d^4 p_1}{(2 \pi)^4} \dfrac{d^4 p_2}{(2 \pi)^4} \ \dfrac{p^a_1 p_{2a}}{p^2_2 (p_1 + p_2)^2}  \nonumber \\
    &(18) \int \dfrac{d^4 p_1}{(2 \pi)^4} \dfrac{d^4 p_2}{(2 \pi)^4} \ \dfrac{p^a_1 p^b_1 p_{2a} p_{2b}}{p^2_1 p^2_2 (p_1 + p_2)^2} \nonumber \\
     &(19) \int \dfrac{d^4 p_1}{(2 \pi)^4} \dfrac{d^4 p_2}{(2 \pi)^4} \ \dfrac{p^a_1 p_{2a}}{p^2_1 p^2_2} \nonumber \\
     &(20) \int \dfrac{d^4 p_1}{(2 \pi)^4} \dfrac{d^4 p_2}{(2 \pi)^4} \ \dfrac{p^a_1 p^b_1 p_{2a} p_{2b}}{p^2_1 p^4_2} \hspace{1cm}  \int \dfrac{d^4 p_1}{(2 \pi)^4} \dfrac{d^4 p_2}{(2 \pi)^4} \ \dfrac{p^a_1 p^b_1 p_{2a} p_{2b}}{p^4_1 p^2_2} \nonumber \\
     &(21) \int \dfrac{d^4 p_1}{(2 \pi)^4} \dfrac{d^4 p_2}{(2 \pi)^4} \ \dfrac{p^a_1 p^b_1 p^c_1 p_{2a} p_{2b} p_{2c}}{p^4_1 p^4_2}.
\end{align}

\noindent
The integrals listed above have been taken directly from MATHEMATICA, where the small Latin indices are the usual space-time indices (for example, $a =\mu$, $b= \nu$, etc.). This is to be noted wherever the loop integrals have been mentioned. To solve these integrals, we use the Veltman-Passarino Reduction method described in \cite{bardin1999standard} to evaluate the divergent part of these integrals within the scheme of dimensional regularization. Appendix \ref{AppC} contains the divergences in all the integrals appearing in this work. Using the Veltman-Passarino Reduction, it is shown in Appendix \ref{AppC} that only non-zero, divergent loop integrals are (4), (11), (12), and (14) in \eqref{45}.  Notice that all the divergent loop integrals have a physical scale; here, we have the mass of the scalar field. Thus diagrams (a), (d), and (e) in Fig.\ref{fig:4.1} will not contribute to the divergent part; only (b) and (c) in Fig.\ref{fig:4.1} will give the divergent part.

\noindent
Now we will calculate,

\begin{align}\label{46}
    I_2 &= \dfrac{i}{2} \Biggl \langle \Biggl( \dfrac{1}{3!} \Bar{S}_{;ijk} \eta^i \eta^j \eta^k +  \dfrac{1}{3!} \Bar{S}_{\text{GF};ijk} \eta^i \eta^j \eta^k + \Bar{S}_{;ij} \eta^i \sigma^j_{(2)} + \Bar{S}_{\text{GF};ij} \eta^i \sigma^j_{(2)} +  \Bar{c}^\alpha \Bar{Q}_{\alpha\beta;i} \eta^i c^\beta \nonumber \\& - \Gamma^{(1)}_{;i} \eta^i \Biggr)^2 \Biggr \rangle.
\end{align}

\noindent
Not all the terms in the above expression correspond to 1PI diagrams. Any term involving $\Gamma^{(1)}_{;i} \eta^i$ would correspond to one particle reducible diagram. For example, if we expand $\langle S_{;ijk}\Gamma^{(1)}_{;l} \eta^i\eta^j\eta^k\eta^l\rangle$ by summing over the Latin indices, then a typical term in the sum would read,

\begin{align}\label{47}
   \int d^4x f(x)\int d^4x' g(x') \langle A(x)B(x)C(x)D(x')\rangle,
\end{align}

\noindent
where we assume,

\begin{align}\label{48}
    \Gamma^{(1)}_{;i} = g(x),
\end{align}

\noindent
with space-time indices suppressed if involved. Note that $\Gamma^{(1)}_{;i}$ corresponds to a one-loop diagram with one vertex.
On using Wick's theorem, we find,

\begin{align}\label{49}
   \int d^4x f(x)\int d^4x' g(x') \langle A(x)B(x)C(x)D(x')\rangle = \int d^4x f(x)\int d^4x' &\Big(\langle A(x)B(x)\rangle\langle C(x)D(x')\rangle \nonumber\\& \hspace{-9cm} +\langle A(x)C(x)\rangle\langle B(x)D(x')\rangle +\langle B(x)C(x)\rangle\langle A(x)D(x')\rangle \Big)g(x').
\end{align}

\noindent
It is clear that each of the terms in the above expression corresponds to one particle reducible diagram. The same argument holds for all terms that involve one factor of $\Bar{c}^\alpha \Bar{Q}_{\alpha\beta;i} \eta^i c^\beta$. However the square of $\Bar{c}^\alpha \Bar{Q}_{\alpha\beta;i} \eta^i c^\beta$ does involve 1PI diagrams. In view of the arguments made above, the expression for $I_2$ simplifies to the following,

\begin{align}\label{50}
    I_2 &= \dfrac{i}{2} \Biggl \langle \Biggl( \dfrac{1}{3!} \Bar{S}_{;ijk} \eta^i \eta^j \eta^k +  \dfrac{1}{3!} \Bar{S}_{\text{GF};ijk} \eta^i \eta^j \eta^k+ \Bar{S}_{;ij} \eta^i \sigma^j_{(2)} + \Bar{S}_{\text{GF};ij} \eta^i \sigma^j_{(2)}\Biggr)^2 \nonumber \\& \hspace{60mm} +  \Biggr(\Bar{c}^\alpha \Bar{Q}_{\alpha\beta;i} \eta^i c^\beta   \Biggr)^2 \Biggr \rangle.
\end{align}

\begin{figure}[h!]
    \centering
    \subfloat[\centering ]
    {\begin{tikzpicture}
    \begin{feynman}
    \node [dot,label=left:\(x\)] (a) at (0,0);
    \vertex (b) at (3,0);
    \node [dot,label=right:\(x'\)] (b) at (3,0);
    \diagram*{
    (a) -- [red, gluon, half left] (b) -- [red, gluon, half left] (a) -- [red, gluon] (b)
    };
    \end{feynman}
    \end{tikzpicture}}
    \qquad
    \subfloat[\centering ]
    {\begin{tikzpicture}
    \begin{feynman}
    \node [dot,label=left:\(x\)] (a) at (0,0);
    \vertex (b) at (3,0);
    \node [dot,label=right:\(x'\)] (b) at (3
    ,0);
    \diagram*{
    (a) -- [blue, half left] (b) -- [blue, half left] (a) -- [blue] (b)
    };
    \end{feynman}
    \end{tikzpicture}}
    \qquad
    \subfloat[\centering ]
    {\begin{tikzpicture}
    \begin{feynman}
    \node [dot,label=left:\(x\)] (a) at (0,0);
    \vertex (b) at (3,0);
    \node [dot,label=right:\(x'\)] (b) at (3,0);
    \diagram*{
    (a) -- [red, gluon, half left] (b) -- [blue, half left] (a) -- [red, gluon] (b)
    };
    \end{feynman}
    \end{tikzpicture}}
    \qquad
    \subfloat[\centering ]
    {\begin{tikzpicture}
    \begin{feynman}
    \node [dot,label=left:\(x\)] (a) at (0,0);
    \vertex (b) at (3,0);
    \node [dot,label=right:\(x'\)] (b) at (3,0);
    \diagram*{
    (a) -- [blue, half left] (b) -- [red, gluon, half left] (a) -- [blue] (b)
    };
    \end{feynman}
    \end{tikzpicture}}
    \qquad
    \subfloat[\centering ]
    {\begin{tikzpicture}
    \begin{feynman}
    \node [dot,label=left:\(x\)] (a) at (0,0);
    \vertex (b) at (3,0);
    \node [dot,label=right:\(x'\)] (b) at (3,0);
    \diagram*{
    (a) -- [red, gluon, half left] (b) -- [scalar, half left] (a) -- [red, gluon] (b)
    };
    \end{feynman}
    \end{tikzpicture}}
    \qquad
    \subfloat[\centering ]
    {\begin{tikzpicture}
    \begin{feynman}
    \node [dot,label=left:\(x\)] (a) at (0,0);
    \vertex (b) at (3,0);
    \node [dot,label=right:\(x'\)] (b) at (3,0);
    \diagram*{
    (a) -- [blue, half left] (b) -- [scalar, half left] (a) -- [blue] (b)
    };
    \end{feynman}
    \end{tikzpicture}}
    
    \caption{\label{fig:4.3} Loop diagrams involved in $I_2$}
\end{figure}
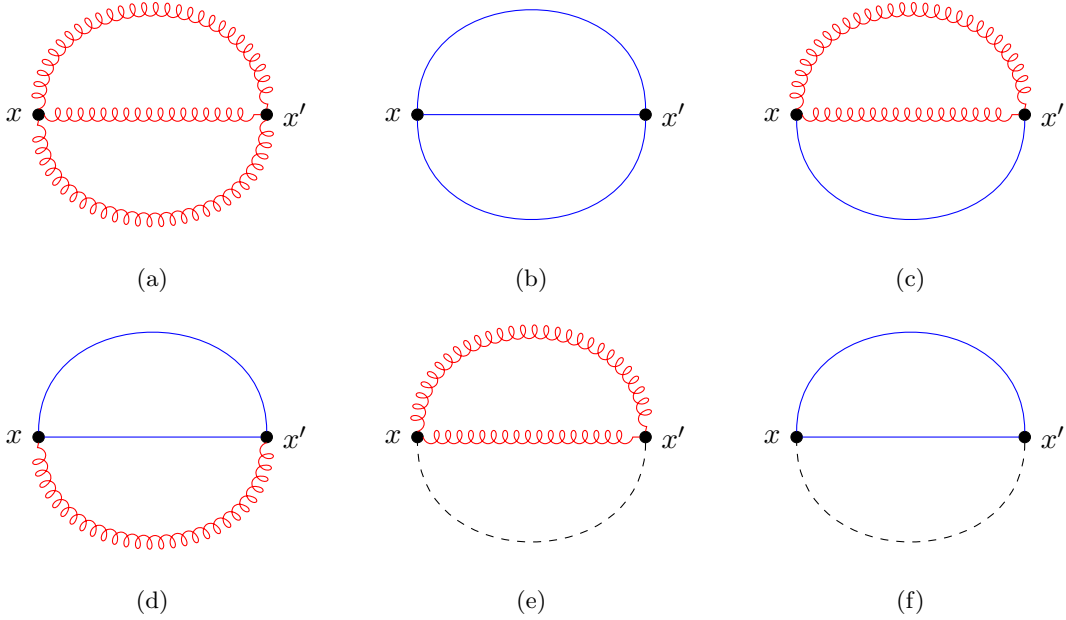

\noindent
On expanding \eqref{50} we get the following six-point functions: $\Bigl \langle h(x) h(x) h(x) h(x') h(x') h(x') \Bigr \rangle$, $\Bigl \langle \psi(x) \psi(x) \psi(x) \psi(x') \psi(x') \psi(x') \Bigr \rangle$, $\Bigl \langle h(x) h(x) h(x') h(x') \psi(x) \psi(x') \Bigr \rangle$, \\ $\Bigl \langle h(x) h(x') \psi(x) \psi(x) \psi(x') \psi(x') \Bigr \rangle$, $\Bigl \langle h(x) h(x) h(x') h(x')\Bar{c}(x) c(x') \Bigr \rangle$, and \\ $\Bigl \langle \psi(x) \psi(x) \psi(x') \psi(x') \Bar{c}(x) c(x')\Bigr \rangle$. which correspond to the diagrams given in Fig.\ref{fig:4.3} respectively. In Fig:\ref{fig:4.3} blue line represents the scalar field propagator, the red spiral represents the graviton propagator, and the black dashed line represents the ghost field propagator.

\noindent
A typical term in the expression for $I_2$ \eqref{50} when expanded, reads,

\begin{align}\label{51}
    \int d^4 x \int d^4 x' f(x) g(x') \Bigl \langle A(x) B(x) C(x) D(x') E(x') F(x') \Bigr \rangle.
\end{align}

\noindent
where as mentioned earlier, $f(x)$ and $g(x)$ are functions of background scalar field and $A(x)$, $B(x)$, ... ,$F(x)$ are placeholders for any of the quantum fields $\{h_{\mu\nu}, \psi, c_\alpha, \bar c_\alpha \}$.

\noindent
Applying Wick's theorem breaks down \eqref{51} into a sum of products of two-point functions,

\begin{align}\label{52}
    &\int d^4 x f(x) \int d^4 x' \int \dfrac{d^4 k}{(2 \pi)^4} \dfrac{d^4 k_1}{(2 \pi)^4} \dfrac{d^4 k_2}{(2 \pi)^4} \dfrac{d^4 k_3}{(2 \pi)^4} e^{i k x} e^{- i (k_1 + k_2 + k_3 - k) x'} \Tilde{g}(k) \nonumber\\&\hspace{30mm}\times\Big(\Bigl \langle A(x) D(x') \Bigr \rangle \Bigl \langle B(x) E(x') \Bigr \rangle   \Bigl \langle C(x) F(x') \Bigr \rangle + \text{all permutations}\Big) \nonumber \\
    &\int d^4 x f(x) \int \dfrac{d^4 k}{(2 \pi)^4} \dfrac{d^4 k_1}{(2 \pi)^4} \dfrac{d^4 k_2}{(2 \pi)^4} \dfrac{d^4 k_3}{(2 \pi)^4} e^{i k x} (2 \pi)^4 \delta^{(4)}(k - (k_1 + k_2 + k_3)) \Tilde{g}(k) \nonumber\\&\hspace{30mm}\times\Big(G_A(k_1) G_B(k_2)   G_C(k_3) + \text{all permutations}\Big) \nonumber \\
    &\int d^4 x f(x) \int \dfrac{d^4 k_1}{(2 \pi)^4} \dfrac{d^4 k_2}{(2 \pi)^4} \dfrac{d^4 k_3}{(2 \pi)^4} e^{i (k_1 + k_2 + k_3) x} \Tilde{g}(k_1 + k_2 + k_3) \nonumber\\&\hspace{30mm}\times\Big(G_A(k_1) G_B(k_2)   G_C(k_3) + \text{all permutations}\Big),
\end{align}

\noindent
where $\tilde g(k)$ is the Fourier transform of $g(x')$,

\begin{equation}\label{53}
    g(x') = \int \dfrac{d^4 k}{(2 \pi)^4} \Tilde{g}(k) e^{i k x'}.
\end{equation}

\noindent
Let's take $q = k_1 + k_2 + k_3$ and $k_3 = q - k_1 - k_2$, use it in \eqref{52},

\begin{align}\label{54}
    &\int d^4 x f(x) \int \dfrac{d^4 q}{(2 \pi)^4} e^{i q x} \Tilde{g}(q) \int \dfrac{d^4 k_1}{(2 \pi)^4} \dfrac{d^4 k_2}{(2 \pi)^4} \Big(G_A(k_1) G_B(k_2) G_C(q-k_1-k_2) + \nonumber \\& \text{all permutations}\Big).
\end{align}

\noindent
We ignore any term that involve factors of $q$ after integrating over $k_1$ and $k_2$, because for any polynomial $F(q)$ in $q$, we find that,

\begin{equation}\label{55}
   \int \dfrac{d^4 q}{(2 \pi)^4} e^{i q x} \Tilde{g}(q) F(q) = F(\partial) g(x).
\end{equation}

\noindent
vanishes if $g(x)$ is a function of a constant background scalar field. Keeping this in mind, we can calculate \eqref{50}. We get the following loop integrals,

\begin{align}\label{56}
    &(1) \int \dfrac{d^4 p_1}{(2 \pi)^4} \dfrac{d^4 p_2}{(2 \pi)^4} \ \dfrac{1}{p^2_1 p^2_2 ((p_1 + p_2 - q)^2 + m^2)}  \nonumber \\
    &(2) \int \dfrac{d^4 p_1}{(2 \pi)^4} \dfrac{d^4 p_2}{(2 \pi)^4} \ \dfrac{1}{p^2_1 (p^2_2 + m^2) ((p_1 + p_2 - q)^2 + m^2)}  \nonumber \\
    &(3) \int \dfrac{d^4 p_1}{(2 \pi)^4} \dfrac{d^4 p_2}{(2 \pi)^4} \ \dfrac{1}{(p^2_1 + m^2) (p^2_2 + m^2) ((p_1 + p_2 - q)^2 + m^2)}  \nonumber \\
    &(4) \int \dfrac{d^4 p_1}{(2 \pi)^4} \dfrac{d^4 p_2}{(2 \pi)^4} \ \dfrac{p^a_1 p^b_1 p_{2a} p_{2b}}{p^4_1 p^4_2 ((p_1 + p_2 - q)^2 + m^2)}  \nonumber \\
    &(5) \int \dfrac{d^4 p_1}{(2 \pi)^4} \dfrac{d^4 p_2}{(2 \pi)^4} \ \dfrac{p^a_1 p^b_1 (q - p_1 - p_2)_{a} (q - p_1 - p_2)_{b}}{p^2_1 (p^2_1 + m^2) p^2_2 (p_1 + p_2 - q)^4}  \nonumber \\
    &(6) \int \dfrac{d^4 p_1}{(2 \pi)^4} \dfrac{d^4 p_2}{(2 \pi)^4} \ \dfrac{p^a_1 p^b_1 p_{2a} p^c_2 (q - p_1 - p_2)_{b} (q - p_1 - p_2)_{c}}{p^2_1 (p^2_1 + m^2) p^4_2 (p_1 + p_2 - q)^4}  \nonumber \\
    &(7) \int \dfrac{d^4 p_1}{(2 \pi)^4} \dfrac{d^4 p_2}{(2 \pi)^4} \ \dfrac{p^a_1 p^b_1 p^c_1 p^d_1 p_{2a} p_{2b} (q - p_1 - p_2)_{c} (q - p_1 - p_2)_{d}}{p^4_1 (p^2_1 + m^2) p^4_2 (p_1 + p_2 - q)^4}  \nonumber \\
    &(8) \int \dfrac{d^4 p_1}{(2 \pi)^4} \dfrac{d^4 p_2}{(2 \pi)^4} \ \dfrac{(q - p_1 - p_2)^2}{(p^2_1 + m^2) (p^2_2 + m^2)}  \nonumber \\
    &(9) \int \dfrac{d^4 p_1}{(2 \pi)^4} \dfrac{d^4 p_2}{(2 \pi)^4} \ \dfrac{p^a_1 p^b_1 p_{2a} p_{2b}}{p^2_1 p^4_2}  \nonumber \\
    &(10) \int \dfrac{d^4 p_1}{(2 \pi)^4} \dfrac{d^4 p_2}{(2 \pi)^4} \ \dfrac{1}{p^2_1}  \nonumber \\
    &(11) \int \dfrac{d^4 p_1}{(2 \pi)^4} \dfrac{d^4 p_2}{(2 \pi)^4} \ \dfrac{p^a_1 p^b_1 p_{2a} p_{2b} (q - p_1 - p_2)^2}{p^4_1 p^4_2}  \nonumber \\
    &(12) \int \dfrac{d^4 p_1}{(2 \pi)^4} \dfrac{d^4 p_2}{(2 \pi)^4} \ \dfrac{(q - p_1 - p_2)^a (q - p_1 - p_2)^b p_{2a} p_{2b}}{p^2_1 p^4_2}  \nonumber \\
    &(13) \int \dfrac{d^4 p_1}{(2 \pi)^4} \dfrac{d^4 p_2}{(2 \pi)^4} \ \dfrac{(q - p_1 - p_2)^2}{p^2_1 p^2_2} \nonumber \\
    &(14) \int \dfrac{d^4 p_1}{(2 \pi)^4}  \dfrac{d^4 p_2}{(2 \pi)^4} \ \dfrac{p^a_1 p^b_1 p_{2a} p_{2b}}{p^2_1 p^2_2 (q - p_1 - p_2)^2}  \nonumber \\
    &(15) \int \dfrac{d^4 p_1}{(2 \pi)^4} \dfrac{d^4 p_2}{(2 \pi)^4} \ \dfrac{p^a_1 p^b_1 p_{2a} p^c_2 (q - p_1 - p_2)_b (q - p_1 - p_2)_c}{p^4_1 p^2_2 (q - p_1 - p_2)^2}.
\end{align}

\noindent
In Appendix \ref{AppC}, it is shown that only (1) to (8) loop integrals of \eqref{56} are divergent. Thus only (b), (c), and (d) loop diagrams from Fig.\ref{fig:4.3} contribute to the divergent part.

\noindent
There is another order $\hbar^2$ contribution coming from one-loop effective action,

\begin{equation}\label{57}
     \Gamma^{(1)}[\Bar{\phi}] = - \hbar \ln{det \Bar{Q}_{\alpha\beta}} + \hbar \lim_{\alpha \to 0} \ \dfrac{1}{2} \ \ln{det\left(\Bar{S}^{;i}_{j} + \dfrac{1}{2\alpha} \ \Bar{K}^{i}_{\beta} \Bar{K}^{\beta}_{j} + \hbar \delta S_{,ij} \right)},
\end{equation}

\noindent
where $\delta S$ is the one-loop counter term. Following the same procedure as \eqref{20}, it is easy to show that,

\begin{equation}\label{58}
     \Gamma^{(1)}[\Bar{\phi}] = - \hbar \ln{det \Bar{Q}_{\alpha\beta}} + \hbar \lim_{\alpha \to 0} \ \dfrac{1}{2} \ \ln{det\left(\Bar{S}^{;i}_{j} + \dfrac{1}{2\alpha} \ \Bar{K}^{i}_{\beta} \Bar{K}^{\beta}_{j} \right)} + i \hbar^2 \delta S_{,ij} \Delta^{ij}(x,x'),
\end{equation}

\noindent
where $\Delta(x,x')$ is propagator matrix, given by

\begin{equation}\label{59}
   \Delta^{ij}(x,x') = \begin{pmatrix}
        G_{\mu \nu \rho \sigma}(x,x') & 0 &\\
        0 & G(x,x') & 
    \end{pmatrix}.
\end{equation}

\noindent
Let's calculate $\delta S_{,ij} \Delta^{ij}(x,x')$. On Fourier transforming, we find, 

\begin{align}\label{60}
    i \delta S_{,ij} \Delta^{ij}(x,x') = i \int d^4 x \delta m^2 \int \dfrac{d^4 p}{(2 \pi)^4} \Tilde{G}(p).
\end{align}

\noindent
Now using \eqref{66}, \eqref{26} and (2) of \eqref{C1}, we get

\begin{align}\label{61}
    \text{div}(i \delta S_{,ij} \Delta^{ij}(x,x')) = - \dfrac{5 m^6}{512 \pi^4 \epsilon^2} - \dfrac{5 m^6}{1024 \pi^4 \epsilon} + \dfrac{5 m^6 \gamma_E}{1024 \pi^4 \epsilon} + \dfrac{5 m^6 \ln\Bigl(\dfrac{m^2}{4 \pi \mu^2}\Bigr)}{1024 \pi^4 \epsilon}.
\end{align}

\noindent
Finally, using \eqref{C21} and \eqref{C44} from Appendix \ref{AppC} into loop integrals \eqref{45} and \eqref{56} and combining these with \eqref{61}, we obtain,

\begin{align}\label{62}
    \Gamma^{(2)}_{div} &= - \int d^4 x \Biggl[ \dfrac{1}{512 \pi^4 \epsilon^2} \ \Biggl(m^6 \kappa^2 - \dfrac{149 m^4 \kappa^2 \phi^2}{288} + \dfrac{9239 m^6 \kappa^4 \phi^2}{1152} \Biggr) + \dfrac{1}{1024 \pi^4 \epsilon} \Biggl(12 m^6 \kappa^2 - \nonumber \\& \hspace{5mm} 9 m^6 \kappa^2 \gamma_E - 9 m^6 \kappa^2 \ln\Bigl(\dfrac{m^2}{4 \pi \mu^2}\Bigr) - \dfrac{149}{288} m^4 \kappa^2 \phi^2 + \dfrac{149}{144} m^4 \gamma_E \kappa^2 \phi^2 - \dfrac{14677}{576} m^6 \kappa^4 \phi^2 +  \nonumber \\& \hspace{5mm} \dfrac{10301}{576} m^6 \gamma_E \kappa^4 \phi^2 + \dfrac{149}{144} m^4 \kappa^2 \phi^2 \ln\Bigl(\dfrac{m^2}{4 \pi \mu^2}\Bigr) + \dfrac{10301}{576} m^6 \kappa^4 \phi^2 \ln\Bigl(\dfrac{m^2}{4 \pi \mu^2}\Bigr) \Biggr) \Biggr].
\end{align}

\noindent
 In \eqref{62}, $\epsilon = 4 - D$, $\gamma_E$ is the Euler–Mascheroni constant, and $\mu$ is the mass dimension 1 scale due to dimension regularisation. We also have terms independent of background field $\phi$ in the expression above; these terms suggest that we need to add a constant term in action. This would lead to the renormalization of cosmological constant had we begun with one in theory.

\section{Renormalization and conclusion}\label{SecIV}

This paper is primarily concerned with an explicit calculation of two-loop divergences in VD effective action for theory \eqref{1}. The expression for divergences has been kept only up to quadratic order in the background scalar field to make calculations simple. Additionally, we also assumed a flat Minkowski metric and a constant scalar field as the background to further ease the computations. In view of this, \eqref{62} forms the main result of this paper. For completeness, we would like to end this section by performing mass renormalization. 
\noindent
To this end, we express the bare quantities in action as,

\begin{equation}\label{63}
    m^2_{B} = m^2 + \hbar (\delta m^{(1)})^2 + \hbar^2 (\delta m^{(2)})^2,
\end{equation}

\noindent
where the renormalized mass is represented by $m$ and $\delta m^{(1)}$ represents its corresponding counterterm at one-loop, $\delta m^{(2)}$ represents its corresponding counterterm at two-loop. The counterterm classical action for flat spacetime at one-loop can be expressed as,

\begin{equation}\label{64}
    \delta S^{(1)} = \int d^4 x \dfrac{1}{2} (\delta m^{(1)})^2 \phi^2.
\end{equation}

\noindent
The divergent terms appearing in the one-loop part of effective action are given in the \eqref{22}. Comparing \eqref{22} and \eqref{64}, we get

\begin{equation}\label{65}
    (\delta m^{(1)})^2 = - \dfrac{5 m^4 \kappa^2}{64 \pi^2 \epsilon}.
\end{equation}

\noindent
Note that the calculations throughout this paper have been carried out for a constant background scalar field. In this limit, the mass correction for the minimal coupling case is in agreement with the results found in \cite{aashish2021covariant} and \cite{mackaytoms} in $\lambda \rightarrow 0$ limit.

\noindent
Now, the counter-term classical action for flat space-time at the two-loop level can be expressed as,

\begin{equation}\label{66}
    \delta S^{(2)} = \int d^4 x \dfrac{1}{2} (\delta m^{(2)})^2 \phi^2
\end{equation}

\noindent
The divergent terms appearing in the two-loop part of effective action are given in \eqref{62}. Comparing \eqref{62} and \eqref{66}, we get

\begin{align}\label{67}
    (\delta m^{(2)})^2 &=  \dfrac{1}{256 \pi^4 \epsilon^2} \ \Biggl(\dfrac{9239 m^6 \kappa^4 \phi^2}{1152} - \dfrac{149 m^4 \kappa^2 \phi^2}{288} \Biggr) - \dfrac{1}{512 \pi^4 \epsilon} \Biggl(\dfrac{149}{288} m^4 \kappa^2 \phi^2 - \nonumber \\& \hspace{5mm} \dfrac{149}{144} m^4 \gamma_E \kappa^2 \phi^2  + \dfrac{14677}{576} m^6 \kappa^4 \phi^2 - \dfrac{10301}{576} m^6 \gamma_E \kappa^4 \phi^2 - \dfrac{149}{144} m^4 \kappa^2 \phi^2 \ln\Bigl(\dfrac{m^2}{4 \pi \mu^2}\Bigr) \nonumber \\& \hspace{5mm} -  \dfrac{10301}{576} m^6 \kappa^4 \phi^2 \ln\Bigl(\dfrac{m^2}{4 \pi \mu^2}\Bigr) \Biggr).
\end{align}

\noindent
We could go a step further and include self-interactions of the scalar field in theory. To discuss the renormalization of coupling constant for $\phi^4$ interaction, we must compute the divergences up to the fourth order in the background field, which forms a part of our future plan. It will also be interesting to add a non-minimal coupling term $\xi\phi^2R$ to the action and discuss the renormalization of $\xi$ at the two-loop level.

\acknowledgments
The calculations in this paper have been carried out in MATHEMATICA using the xAct packages xTensor \cite{xtensor}, and xPert \cite{xpert}. It is a pleasure to thank Mr. Archit Vidyarthi for the fruitful discussions. This work is partially supported by the DST (Govt. of India) Grant No. SERB/PHY/2021057.

\appendix
\section{Derivation of Two-Loop Effective Action}\label{AppA}

From \eqref{10} we can write,
 
 \begin{equation}\label{A1}
     e^{\dfrac{i}{\hbar} \ \Gamma_\alpha} = \int \prod_{i} d\sigma^i \sqrt{g^*} [Dc][D\Bar{c}] \  e^{\dfrac{i}{\hbar}\left[ S[\phi_*,\sigma^i] + S_{GF}[\phi_*,\sigma^i] + \Bar{c}^\alpha Q_{\alpha\beta} c^\beta + \dfrac{\delta \Gamma_\alpha}{\delta v^i} \left( v^i - \sigma^i \right) \right]}.
 \end{equation}
 
 \noindent
 Now taylor expand $S$ and $S_{GF}$ around $\sigma^i[\phi_*;\phi] = v^i[\phi_*;\Bar{\phi}]$
 
 \[
     S[\phi_*,\sigma^i[\phi_*;\phi]] = S[\phi_*,v^i] + \sum_{n = 1}^{\infty} \ \dfrac{1}{n!} \ \dfrac{\delta^n S[\phi_*;v^i]}{\delta v^{i_1} \cdots \delta v^{i_n}} \left(\sigma^{i_1} - v^{i_1}\right) \cdots  \left(\sigma^{i_n} - v^{i_n}\right),
 \]
 
 \[
     S_{GF}[\phi_*,\sigma^i[\phi_*;\phi]] = S_{GF}[\phi_*,v^i] + \sum_{n = 1}^{\infty} \ \dfrac{1}{n!} \ \dfrac{\delta^n S_{GF}[\phi_*;v^i]}{\delta v^{i_1} \cdots \delta v^{i_n}} \left(\sigma^{i_1} - v^{i_1}\right) \cdots  \left(\sigma^{i_n} - v^{i_n}\right),
 \]
 
 \[
     Q[\phi_*,\sigma^i[\phi_*;\phi]] = Q[\phi_*,v^i] + \sum_{n = 1}^{\infty} \ \dfrac{1}{n!} \ \dfrac{\delta^n Q[\phi_*;v^i]}{\delta v^{i_1} \cdots \delta v^{i_n}} \left(\sigma^{i_1} - v^{i_1}\right) \cdots  \left(\sigma^{i_n} - v^{i_n}\right).
 \]
 
 \noindent
 Using Covariant Taylor expansion, one can write,
 
 \[\dfrac{\delta^n S[\phi_*;v^i]}{\delta v^{i_1} \cdots \delta v^{i_n}} = (-1)^n S_{;(i_1 \cdots i_n)}[\Bar{\phi}].\]
 
 \begin{equation}\label{A2}
     S[\phi_*,\sigma^i[\phi_*;\phi]] = S[\phi_*,v^i] + \sum_{n = 1}^{\infty} \ \dfrac{(-1)^n}{n!} \ S_{;(i_1 \cdots i_n)}[\Bar{\phi}] \left(\sigma^{i_1} - v^{i_1}\right) \cdots  \left(\sigma^{i_n} - v^{i_n}\right),
 \end{equation}

\begin{equation}\label{A3}
     S_{GF}[\phi_*,\sigma^i[\phi_*;\phi]] = S_{GF}[\phi_*,v^i] + \sum_{n = 1}^{\infty} \ \dfrac{(-1)^n}{n!} \ S_{GF;(i_1 \cdots i_n)}[\Bar{\phi}] \left(\sigma^{i_1} - v^{i_1}\right) \cdots  \left(\sigma^{i_n} - v^{i_n}\right),
 \end{equation}

\begin{equation}\label{A4}
     Q[\phi_*,\sigma^i[\phi_*;\phi]] = Q[\phi_*,v^i] + \sum_{n = 1}^{\infty} \ \dfrac{(-1)^n}{n!} \ Q_{;(i_1 \cdots i_n)}[\Bar{\phi}] \left(\sigma^{i_1} - v^{i_1}\right) \cdots  \left(\sigma^{i_n} - v^{i_n}\right).
 \end{equation}

\noindent
Let's redefine variable $\sigma^i - v^i = \Tilde{\sigma}^i$. Using all the above results in eq. \eqref{A1} and also for the time-being ignore the $\alpha$ subscript in the same equation,

 \begin{align}\label{A5}
     e^{\dfrac{i}{\hbar} \ \Gamma} &= \int \prod_{i} d\Tilde{\sigma}^i \sqrt{g^*} [Dc][D\Bar{c}] \ exp \Bigl[ \dfrac{i}{\hbar}\Big\{\Bar{S} + \Bar{S}_{;i} \Tilde{\sigma}^i + \dfrac{1}{2!} \Bar{S}_{;ij} \Tilde{\sigma}^i \Tilde{\sigma}^j + \dfrac{1}{3!} \Bar{S}_{;ijk} \Tilde{\sigma}^i \Tilde{\sigma}^j \Tilde{\sigma}^k + \nonumber \\& \dfrac{1}{4!} \Bar{S}_{;ijkl} \Tilde{\sigma}^i \Tilde{\sigma}^j \Tilde{\sigma}^k \Tilde{\sigma}^l  + \mathcal{O}( \Tilde{\sigma}^5) +  \Bar{S}_{GF} + \Bar{S}_{GF;i} \Tilde{\sigma}^i + \dfrac{1}{2!} \Bar{S}_{GF;ij} \Tilde{\sigma}^i \Tilde{\sigma}^j + \dfrac{1}{3!} \Bar{S}_{GF;ijk} \Tilde{\sigma}^i \Tilde{\sigma}^j \Tilde{\sigma}^k + \nonumber \\& \dfrac{1}{4!} \Bar{S}_{GF;ijkl} \Tilde{\sigma}^i \Tilde{\sigma}^j \Tilde{\sigma}^k \Tilde{\sigma}^l  + \mathcal{O}( \Tilde{\sigma}^5) +  \Bar{C}^\alpha \Bar{Q}_{\alpha \beta} C^\beta + \Bar{C}^\alpha \Bar{Q}_{\alpha \beta;i} C^\beta \Tilde{\sigma}^i + \dfrac{1}{2!} \Bar{C}^\alpha \Bar{Q}_{\alpha \beta;ij} C^\beta \Tilde{\sigma}^i \Tilde{\sigma}^j + \nonumber \\& \mathcal{O}( \Tilde{\sigma}^3) - \Gamma_{,i} \Tilde{\sigma}^i \Bigr\}. \Bigr]
 \end{align}

\noindent
In \eqref{A5}, the expansion involves the covariant Taylor expansion of the classical action in powers of $\sigma^i$, and we must now use the full connection ($\Tilde{\Gamma}^i_{jk}$) to calculate the various covariant derivatives that are in the expansion. Note that the classical action functional is most naturally expressed in terms of $\phi^i$ rather than $\sigma^i$; we must look at how $\sigma^i$ depends on $\phi^i$. In general, this dependence is very complicated, but in the Landau–DeWitt gauge, it is a much simpler result. We can Taylor expand $\sigma^i$ about $\phi = \phi_*$,

\[\sigma^i[\phi_*;\phi] = - \eta^i + \sum_{n = 2}^{\infty} \sigma^i_{(n)}, \]

\noindent
where

\[\sigma^i_{(n)} = \dfrac{1}{n!} \sigma^i_{j_1 \cdots j_n} \eta^{j_1} \cdots \eta^{j_n}. \]

\noindent
For our case we will keep terms up to $\sigma_{(4)}$,

\begin{equation}\label{A6}
    \sigma^i = - \eta^i + \sigma^i_{(2)} + \sigma^i_{(3)} + \sigma^i_{(4)}.
\end{equation}

\noindent
Use the \eqref{A6} in the \eqref{A5} and reparametrize the fields as $\eta^i \xrightarrow{} \sqrt{\hbar} \eta^i$, $\Bar{C}^\alpha \xrightarrow{} \sqrt{\hbar} \Bar{C}^\alpha$ and $C^\beta \xrightarrow{}\sqrt{\hbar} C^\beta$. Putting all this together in \eqref{A5} and simplifying we get,

\begin{multline}\label{A7}
     e^{\dfrac{i}{\hbar} \ \Gamma} = \int \prod_{i} d\Tilde{\sigma}^i \sqrt{g^*} [Dc][D\Bar{c}] \ exp \Bigl(i \Big\{\dfrac{1}{\hbar}\Bigl[\Bar{S} + \Bar{S} \Bigr] + \Bigl[\dfrac{1}{2!} \Bar{S}_{;ij} \eta^i \eta^j + \dfrac{1}{2!} \Bar{S}_{GF;ij} \eta^i \eta^j + \Bar{C}^\alpha \Bar{Q}_{\alpha \beta} C^\beta \Bigr] \\ - \sqrt{\hbar} \Bigl[\dfrac{1}{3!} \Bar{S}_{;ijk} \eta^i \eta^j \eta^k + \dfrac{1}{3!} \Bar{S}_{GF;ijk} \eta^i \eta^j \eta^k + \Bar{S}_{;ij} \eta^i \sigma^i_{(2)} + \Bar{S}_{GF;ij} \eta^i \sigma^i_{(2)}  + \Bar{C}^\alpha \Bar{Q}_{\alpha \beta;i} C^\beta \eta^i - \Gamma^{(1)}_{;i} \eta^i \Bigr] + \\  \hbar\Bigl[\dfrac{1}{4!} \Bar{S}_{;ijkl} \eta^i \eta^j \eta^k \eta^l + \dfrac{1}{4!} \Bar{S}_{GF;ijkl} \eta^i \eta^j \eta^k \eta^l + \dfrac{1}{2!} \Bar{S}_{;ijk} \eta^i \eta^j \sigma^k_{(2)} + \dfrac{1}{2!} \Bar{S}_{GF;ijk} \eta^i \eta^j \sigma^k_{(2)} + \dfrac{1}{2!} \Bar{S}_{;ij} \sigma^i_{(2)}  \sigma^j_{(2)} + \\ \dfrac{1}{2!} \Bar{S}_{GF;ij} \sigma^i_{(2)}  \sigma^j_{(2)} - \Bar{S}_{;ij} \eta^i \sigma^j_{(3)} - \Bar{S}_{;ij} \eta^i \sigma^j_{(3)} + \Bar{C}^\alpha \Bar{Q}_{\alpha \beta;i} C^\beta \sigma^i_{(2)} + \dfrac{1}{2!} \Bar{C}^\alpha \Bar{Q}_{\alpha \beta;ij} C^\beta \eta^i \eta^j - \Gamma^{(1)}_{;i} \sigma^i_{(2)} \Bigr] \Bigr\} \Bigl),
 \end{multline}

\noindent
where,

\begin{equation}\label{A8}
    \sigma^i_{(2)} = -\dfrac{1}{2} \Tilde{\Gamma}^i_{jk} \eta^j \eta^k,
\end{equation}

\begin{equation}\label{A9}
    \sigma^i_{(3)} = -\dfrac{1}{3!} \Bigl(\Tilde{\Gamma}^i_{jk,l} + \Tilde{\Gamma}^i_{jm} \Tilde{\Gamma}^m_{kl} \Bigr) \eta^j \eta^k \eta^l,
\end{equation}

 \noindent
 and,
 
\begin{equation}\label{A10}
    \Tilde{\Gamma}^i_{jk} = \Gamma^i_{jk} + \dfrac{1}{2} \gamma^{\alpha\epsilon} \gamma^{\beta\sigma} K_{\alpha j} K_{\beta k} \Bigl( K^n_\epsilon K^i_{\sigma;n} + K^n_\sigma K^i_{\epsilon;n} \Bigr) - \gamma^{\alpha \beta} \Bigl( K_{\alpha j} K^i_{\beta;k} + K_{\alpha k} K^i_{\beta;j} \Bigr).
\end{equation}

\noindent
Note that the measure part will also transform under transformation \eqref{A6} as $d\sigma^i = |J| d\eta^i$, here $|J|$ is the Jacobian matrix of the transformation. Here we will keep terms up to $\sigma^i_{(2)}$ because higher terms will not contribute in two loops. Remember that we have two fields $\Bigl\{h_{\mu \nu}, \phi \Bigr\}$; in general, for n fields calculating the Jacobian is highly non-trivial and complicated. But, If we use the Faddeev–Popov method for this jacobian, we see that at the two-loop level, we can write $d\sigma^i = |J| d\eta^i$

\begin{equation}\label{A11}
    J^i_l = \dfrac{d\sigma^i}{d\eta^l} = -\sqrt{\hbar} \delta^i_l - \hbar  \Tilde{\Gamma}^i_{lk} \eta^k.
\end{equation}

\noindent
Due to the presence of $\hbar$ in \eqref{A11}, the Jacobian term gives $\hbar^{3/2}$ as the lowest order term. Thus this term will not contribute at the two-loop level. Thus,

\begin{align}\label{A12}
     e^{\dfrac{i}{\hbar} \ \Gamma} &= \int \prod_{i} d\eta^i \sqrt{g^*} [Dc][D\Bar{c}] \ exp \Bigl(i \Big\{\dfrac{1}{\hbar}\Bigl[\Bar{S} + \Bar{S} \Bigr] + \Bigl[\dfrac{1}{2!} \Bar{S}_{;ij} \eta^i \eta^j  + \dfrac{1}{2!} \Bar{S}_{GF;ij} \eta^i \eta^j + \Bar{C}^\alpha \Bar{Q}_{\alpha \beta} C^\beta \Bigr] \nonumber \\& - \sqrt{\hbar} \Bigl[\dfrac{1}{3!} \Bar{S}_{;ijk} \eta^i \eta^j \eta^k +  \dfrac{1}{3!} \Bar{S}_{GF;ijk} \eta^i \eta^j \eta^k + \Bar{S}_{;ij} \eta^i \sigma^i_{(2)} +  \Bar{S}_{GF;ij} \eta^i \sigma^i_{(2)}  + \Bar{C}^\alpha \Bar{Q}_{\alpha \beta;i} C^\beta \eta^i - \nonumber \\& \Gamma^{(1)}_{;i} \eta^i \Bigr] +  \hbar\Bigl[\dfrac{1}{4!} \Bar{S}_{;ijkl} \eta^i \eta^j \eta^k \eta^l + \dfrac{1}{4!} \Bar{S}_{GF;ijkl} \eta^i \eta^j \eta^k \eta^l + \dfrac{1}{2!} \Bar{S}_{;ijk} \eta^i \eta^j \sigma^k_{(2)}  + \dfrac{1}{2!} \Bar{S}_{GF;ijk} \eta^i \eta^j \sigma^k_{(2)} \nonumber \\& + \dfrac{1}{2!} \Bar{S}_{;ij} \sigma^i_{(2)}  \sigma^j_{(2)} +  \dfrac{1}{2!} \Bar{S}_{GF;ij} \sigma^i_{(2)}  \sigma^j_{(2)} - \Bar{S}_{;ij} \eta^i \sigma^j_{(3)} - \Bar{S}_{;ij} \eta^i \sigma^j_{(3)} + \Bar{C}^\alpha \Bar{Q}_{\alpha \beta;i} C^\beta \sigma^i_{(2)} + \nonumber \\& \dfrac{1}{2!} \Bar{C}^\alpha \Bar{Q}_{\alpha \beta;ij} C^\beta \eta^i \eta^j -  \Gamma^{(1)}_{;i} \sigma^i_{(2)} \Bigr] \Bigr\} \Bigl).
 \end{align}

\noindent
For simplicity, let's define the following,

\begin{align*}
    &S^{(0)} = \Bar{S} + \Bar{S}_{GF}, \nonumber \\
    &S^{(2)} = \dfrac{1}{2!} \Bar{S}_{;ij} \eta^i \eta^j  + \dfrac{1}{2!} \Bar{S}_{GF;ij} \eta^i \eta^j + \Bar{C}^\alpha \Bar{Q}_{\alpha \beta} C^\beta, \nonumber \\
    &A_1 = \dfrac{1}{3!} \Bar{S}_{;ijk} \eta^i \eta^j \eta^k +  \dfrac{1}{3!} \Bar{S}_{GF;ijk} \eta^i \eta^j \eta^k + \Bar{S}_{;ij} \eta^i \sigma^i_{(2)} +  \Bar{S}_{GF;ij} \eta^i \sigma^i_{(2)}  + \Bar{C}^\alpha \Bar{Q}_{\alpha \beta;i} C^\beta \eta^i - \Gamma^{(1)}_{;i} \eta^i, \nonumber \\
    &A_2 = \dfrac{1}{4!} \Bar{S}_{;ijkl} \eta^i \eta^j \eta^k \eta^l + \dfrac{1}{4!} \Bar{S}_{GF;ijkl} \eta^i \eta^j \eta^k \eta^l + \dfrac{1}{2!} \Bar{S}_{;ijk} \eta^i \eta^j \sigma^k_{(2)} +  \dfrac{1}{2!} \Bar{S}_{GF;ijk} \eta^i \eta^j \sigma^k_{(2)} + \nonumber \\& \dfrac{1}{2!} \Bar{S}_{;ij} \sigma^i_{(2)}  \sigma^j_{(2)} + \dfrac{1}{2!} \Bar{S}_{GF;ij} \sigma^i_{(2)}  \sigma^j_{(2)} - \Bar{S}_{;ij} \eta^i \sigma^j_{(3)} - \Bar{S}_{;ij} \eta^i \sigma^j_{(3)} + \Bar{C}^\alpha \Bar{Q}_{\alpha \beta;i} C^\beta \sigma^i_{(2)} + \nonumber \\& \dfrac{1}{2!} \Bar{C}^\alpha \Bar{Q}_{\alpha \beta;ij} C^\beta \eta^i \eta^j -  \Gamma^{(1)}_{;i} \sigma^i_{(2)}.
\end{align*}

\begin{multline}\label{A13}
     e^{\dfrac{i}{\hbar} \ \Gamma} = \int \prod_{i} d\eta^i \sqrt{g^*} [Dc][D\Bar{c}] \ exp \Bigl(i \Big\{\dfrac{1}{\hbar} S^{(0)} + S^{(2)} - \sqrt{\hbar} A_1 +  \hbar A_2 \Bigr\} \Bigl).
 \end{multline}

\noindent
Let's focus on exponential tern in \eqref{A13}. Since $S^{(0)}$ term is functional of the background field, we can take it outside of the integral. We will keep terms up to $\hbar$ order in the expansion.

\[\Rightarrow \ e^{\dfrac{i}{\hbar} S^{(0)}}\int \prod_{i} d\eta^i \sqrt{g^*} [Dc][D\Bar{c}] \ e^{i S^{(2)}} e^{i \Big\{ - \sqrt{\hbar} A_1 +  \hbar A_2 \Bigr\} }, \]

\[\Rightarrow \ e^{\dfrac{i}{\hbar} S^{(0)}}\int \prod_{i} d\eta^i \sqrt{g^*} [Dc][D\Bar{c}] \ e^{i S^{(2)}} \Bigl[1  - i \sqrt{\hbar} A_1 + i \hbar A_2  - \dfrac{1}{2} \Bigl(- \sqrt{\hbar} A_1 +  \hbar A_2\Bigr)^2 + \cdots\Bigr],\]

\begin{multline*}
    \Rightarrow \ e^{\dfrac{i}{\hbar} S^{(0)}}\Biggl[\int \prod_{i} d\eta^i \sqrt{g^*} [Dc][D\Bar{c}] \ e^{i S^{(2)}} -  i \sqrt{\hbar} \int \prod_{i} d\eta^i \sqrt{g^*} [Dc][D\Bar{c}] \ e^{i S^{(2)}} A_1 + \\ i \hbar \int \prod_{i} d\eta^i \sqrt{g^*} [Dc][D\Bar{c}] \ e^{i S^{(1)}} A_2  - \dfrac{1}{2} \hbar \int \prod_{i} d\eta^i \sqrt{g^*} [Dc][D\Bar{c}] \ e^{i S^{(2)}} A^2_1 +\mathcal{O}(\hbar^{3/2})  \Biggr],
\end{multline*}

\begin{equation*}
    \Rightarrow \ e^{\dfrac{i}{\hbar} S^{(0)}}\int \prod_{i} d\eta^i \sqrt{g^*} [Dc][D\Bar{c}] \ e^{i S^{(2)}} \Bigl[1  - i \sqrt{\hbar} \left\langle{A_1}\right\rangle + i \hbar \bigl\langle{A_2}\bigr\rangle  - \dfrac{1}{2} \hbar \bigl\langle{A^2_1}\bigr\rangle \Bigr].
\end{equation*}

\noindent
Where,

\[\bigl\langle{F}\bigr\rangle = \dfrac{\int \prod_{i} d\eta^i \sqrt{g^*} [Dc][D\Bar{c}] \ F \ e^{i S^{(2)}}}{\int \prod_{i} d\eta^i \sqrt{g^*} [Dc][D\Bar{c}] \ e^{i S^{(2)}}}. \]

\noindent
In the above equation, $\left\langle{A_1}\right\rangle$ has odd numbers of $\eta^i$, so the average over the odd point function is zero,

\begin{equation}\label{A14}
    \Rightarrow \ e^{\dfrac{i}{\hbar} S^{(0)}}\int \prod_{i} d\eta^i \sqrt{g^*} [Dc][D\Bar{c}] \ e^{i S^{(2)}} \Bigl[1  + i \hbar \bigl\langle{A_2}\bigr\rangle  - \dfrac{1}{2} \hbar \bigl\langle{A^2_1}\bigr\rangle \Bigr].
\end{equation}

\noindent
Now expand the effective action in the powers of the $\hbar$ on the left hand side of the \eqref{A13} using \eqref{7} upto $n = 2$, we get,

\begin{equation}\label{A15}
    \begin{aligned}
        \Gamma^{(0)}[\Bar{\phi}] &= \Bar{S},
    \end{aligned}
 \end{equation}

 \begin{equation}\label{A16}
     \Gamma^{(1)}[\Bar{\phi}] = - \ln{det \Bar{Q}_{\alpha\beta}} + \lim_{\alpha \to 0} \dfrac{1}{2} \ \ln{det\left(\Bar{S}^{;i}_{j} + \dfrac{1}{2\alpha} \ \Bar{K}^{i}_{\beta} \Bar{K}^{\beta}_{j} \right)},
 \end{equation}

\[\Gamma^{(2)}[\Bar{\phi}] = \bigl\langle{A_2}\bigr\rangle  + \dfrac{i}{2} \bigl\langle{A^2_1}\bigr\rangle, \]

     \begin{align}\label{A17}
        \Gamma^{(2)}[\Bar{\phi}] &=  \Biggl \langle \dfrac{1}{4!} \Bar{S}_{;ijkl} \eta^i \eta^j \eta^k \eta^l + \dfrac{1}{4!} \Bar{S}_{GF;ijkl} \eta^i \eta^j \eta^k \eta^l + \dfrac{1}{2} \Bar{S}_{;ij} \sigma^i_{(2)} \sigma^j_{(2)} + \dfrac{1}{2} \Bar{S}_{GF;ij} \sigma^i_{(2)} \sigma^j_{(2)} + \nonumber \\& \hspace{7mm} \dfrac{1}{2} \Bar{S}_{;ijk} \eta^i \eta^j \sigma^k_{(2)} + \dfrac{1}{2} \Bar{S}_{GF;ijk} \eta^i \eta^j \sigma^k_{(2)} - \Bar{S}_{;ij} \eta^i \sigma^j_{(3)} - \Bar{S}_{GF;ij} \eta^i \sigma^j_{(3)} +  \Bar{c}^\alpha \Bar{Q}_{\alpha\beta;i} \sigma^i_{(2)} c^\beta + \nonumber \\& \hspace{7mm} \dfrac{1}{2} \Bar{c}^\alpha \Bar{Q}_{\alpha \beta;ij} \eta^i \eta^j c^\beta  - \Gamma^{(1)}_{;i} \sigma^i_{(2)} \Biggr \rangle  + \dfrac{i}{2} \Biggl \langle \Biggl( \dfrac{1}{3!} \Bar{S}_{;ijk} \eta^i \eta^j \eta^k +  \dfrac{1}{3!} \Bar{S}_{GF;ijk} \eta^i \eta^j \eta^k + \nonumber \\& \hspace{7mm}\Bar{S}_{;ij} \eta^i \sigma^j_{(2)}  + \Bar{S}_{GF;ij} \eta^i \sigma^j_{(2)} +  \Bar{c}^\alpha \Bar{Q}_{\alpha\beta;i} \eta^i c^\beta  - \Gamma^{(1)}_{;i} \eta^i \Biggr)^2 \Biggr \rangle.
    \end{align}

\section{Simplifications in the Landau-DeWitt gauge}\label{AppB}

\noindent
From VD connection given in \eqref{23}, we define

\begin{align}\label{B1}
    T^i_{jk} &= \dfrac{1}{2} \gamma^{\alpha\epsilon} \gamma^{\beta\sigma} K_{\alpha j} K_{\beta k} \Bigl( K^n_\epsilon K^i_{\sigma;n} + K^n_\sigma K^i_{\epsilon;n} \Bigr) - \gamma^{\alpha \beta} \Bigl( K_{\alpha j} K^i_{\beta;k} + K_{\alpha k} K^i_{\beta;j} \Bigr) \nonumber \\
    &=  K^m_\sigma K^i_{\tau;m} B^\sigma_{(j} B^\tau_{k)} - 2 K^i_{\alpha:(j} B^\alpha_{k)},
\end{align}

\noindent
where $B^\alpha_i = \gamma^{\alpha \beta} K_{i \beta}[\varphi_*] $ and, $A_{(i} B_{j)} = \dfrac{1}{2} (A_i B_j + A_j B_i)$. Since action is scalar, we can write,

\begin{equation}\label{B2}
    S_{;i} = S_{,i}.
\end{equation}

\noindent
From the second covariant derivatives onward, we need to use the Vilkovisky-DeWitt Connections to calculate the covariant derivatives.

\begin{align}\label{B3}
    &S_{;ij} = (S_{,i})_{;j} \nonumber \\
    &S_{;ij} = S_{,ij} - \Tilde{\Gamma}^k_{ij} S_{,k},
\end{align}

\begin{align}\label{B4}
    &S_{;ijk} = (S_{;ij})_{;k} \nonumber \\& \hspace{7.5mm} = (S_{;ij})_{,k} - \Tilde{\Gamma}^m_{ik} S_{;mj} - \Tilde{\Gamma}^m_{jk} S_{;im} \nonumber \\& \hspace{7.5mm} = (S_{,ij} - \Tilde{\Gamma}^n_{ij} S_{,n})_{,k} - \Tilde{\Gamma}^m_{jk} (S_{,im} - \Tilde{\Gamma}^n_{im} S_{,n}) - \Tilde{\Gamma}^m_{ik} (S_{,jm} - \Tilde{\Gamma}^n_{jm} S_{,n}) \nonumber \\
    &S_{;ijk} = S_{,ijk} - \Tilde{\Gamma}^m_{ij,k} S_{,m} - \Tilde{\Gamma}^m_{ij} S_{,mk} - \Tilde{\Gamma}^m_{ik} S_{,jm} - \Tilde{\Gamma}^m_{jk} S_{,im} + \Tilde{\Gamma}^m_{ik} \Tilde{\Gamma}^n_{jm} S_{,n} + \Tilde{\Gamma}^m_{jk} \Tilde{\Gamma}^n_{im} S_{,n},
\end{align}

\begin{align}\label{B5}
    &S_{;ijkl} = (S_{;ijk})_{;l} \nonumber \\& \hspace{8mm} = (S_{;ijk})_{,l} - \Tilde{\Gamma}^m_{il} S_{;mjk} - \Tilde{\Gamma}^m_{jl} S_{;imk} - \Tilde{\Gamma}^m_{kl} S_{;ijm} \nonumber \\& \hspace{8mm} = (S_{,ijk} - \Tilde{\Gamma}^m_{ij,k} S_{,m} - \Tilde{\Gamma}^m_{ij} S_{,mk} - \Tilde{\Gamma}^m_{ik} S_{,jm} - \Tilde{\Gamma}^m_{jk} S_{,im} + \Tilde{\Gamma}^m_{ik} \Tilde{\Gamma}^n_{jm} S_{,n} + \Tilde{\Gamma}^m_{jk} \Tilde{\Gamma}^n_{im} S_{,n})_{,l} - \nonumber \\& \hspace{13mm} \Tilde{\Gamma}^m_{il} (S_{,mjk} - \Tilde{\Gamma}^n_{mj,k} S_{,n} - \Tilde{\Gamma}^n_{mj} S_{,nk} - \Tilde{\Gamma}^n_{mk} S_{,jn} - \Tilde{\Gamma}^n_{jk} S_{,nm} + \Tilde{\Gamma}^n_{mk} \Tilde{\Gamma}^p_{jn} S_{,p} + \Tilde{\Gamma}^n_{jk} \Tilde{\Gamma}^p_{nm} S_{,p}) - \nonumber \\& \hspace{13mm} \Tilde{\Gamma}^m_{jl} (S_{,imk} - \Tilde{\Gamma}^n_{im,k} S_{,n} - \Tilde{\Gamma}^n_{im} S_{,nk} - \Tilde{\Gamma}^n_{ik} S_{,mn} - \Tilde{\Gamma}^n_{mk} S_{,in} + \Tilde{\Gamma}^n_{ik} \Tilde{\Gamma}^p_{mn} S_{,p} + \Tilde{\Gamma}^n_{mk} \Tilde{\Gamma}^p_{in} S_{,p}) - \nonumber \\& \hspace{13mm}\Tilde{\Gamma}^m_{kl} (S_{,ijm} - \Tilde{\Gamma}^n_{ij,m} S_{,n} - \Tilde{\Gamma}^n_{ij} S_{,mn} - \Tilde{\Gamma}^n_{im} S_{,jn} - \Tilde{\Gamma}^n_{jm} S_{,in} + \Tilde{\Gamma}^n_{im} \Tilde{\Gamma}^p_{jn} S_{,p} + \Tilde{\Gamma}^n_{jm} \Tilde{\Gamma}^p_{in} S_{,p}) \nonumber \\
    &S_{;ijkl} = S_{,ijkl} - \Tilde{\Gamma}^m_{ij} S_{,mkl} - \Tilde{\Gamma}^m_{ik} S_{,mjl} - \Tilde{\Gamma}^m_{il} S_{,mjk} - \Tilde{\Gamma}^m_{jk} S_{,iml} - \Tilde{\Gamma}^m_{jl} S_{,imk} - \Tilde{\Gamma}^m_{kl} S_{,ijm} - \Tilde{\Gamma}^m_{ij,k} S_{,ml}  \nonumber \\& \hspace{13mm} - \Tilde{\Gamma}^m_{ij,l} S_{,mk} -  \Tilde{\Gamma}^m_{ik,l} S_{,mj} - \Tilde{\Gamma}^m_{jk,l} S_{,mi} + \Tilde{\Gamma}^m_{ik} \Tilde{\Gamma}^n_{jm} S_{,nl} + \Tilde{\Gamma}^m_{jk} \Tilde{\Gamma}^n_{im} S_{,nl} + \Tilde{\Gamma}^m_{il} \Tilde{\Gamma}^n_{jm} S_{,nk} \nonumber \\&  \hspace{13mm} + \Tilde{\Gamma}^m_{il} \Tilde{\Gamma}^n_{km} S_{,nj}  + \Tilde{\Gamma}^m_{jl} \Tilde{\Gamma}^n_{im} S_{,nk} + \Tilde{\Gamma}^m_{jl} \Tilde{\Gamma}^n_{mk} S_{,ni} + \Tilde{\Gamma}^m_{kl} \Tilde{\Gamma}^n_{im} S_{,nj} + \Tilde{\Gamma}^m_{kl} \Tilde{\Gamma}^n_{jm} S_{,ni} + \Tilde{\Gamma}^m_{kl} \Tilde{\Gamma}^n_{ij} S_{,nm} \nonumber \\& \hspace{13mm} + \Tilde{\Gamma}^m_{jl} \Tilde{\Gamma}^n_{ik} S_{,nm} +  \Tilde{\Gamma}^m_{ij} \Tilde{\Gamma}^n_{kl} S_{,nm} -  \Tilde{\Gamma}^m_{ij,kl} S_{,m} + \Tilde{\Gamma}^m_{ik,l} \Tilde{\Gamma}^n_{jm} S_{,n} + \Tilde{\Gamma}^m_{jk,l} \Tilde{\Gamma}^n_{im} S_{,n} + \Tilde{\Gamma}^m_{ik} \Tilde{\Gamma}^n_{jm,l} S_{,n} \nonumber \\& \hspace{13mm}  + \Tilde{\Gamma}^m_{jk} \Tilde{\Gamma}^n_{im,l} S_{,n} +  \Tilde{\Gamma}^m_{il} \Tilde{\Gamma}^n_{jm,k} S_{,n} + \Tilde{\Gamma}^m_{jl} \Tilde{\Gamma}^n_{im,k} S_{,n} +  \Tilde{\Gamma}^m_{kl} \Tilde{\Gamma}^n_{ij,k} S_{,n} - \Tilde{\Gamma}^m_{il} \Tilde{\Gamma}^n_{mk} \Tilde{\Gamma}^p_{jn} S_{,p} \nonumber \\& \hspace{13mm} - \Tilde{\Gamma}^m_{jl} \Tilde{\Gamma}^n_{mk} \Tilde{\Gamma}^p_{in} S_{,p} -  \Tilde{\Gamma}^m_{kl} \Tilde{\Gamma}^n_{im} \Tilde{\Gamma}^p_{jn} S_{,p} - \Tilde{\Gamma}^m_{kl} \Tilde{\Gamma}^n_{jm} \Tilde{\Gamma}^p_{in} S_{,p} - \Tilde{\Gamma}^m_{il} \Tilde{\Gamma}^n_{jk} \Tilde{\Gamma}^p_{mn} S_{,p} -  \Tilde{\Gamma}^m_{jl} \Tilde{\Gamma}^n_{ik} \Tilde{\Gamma}^p_{mn} S_{,p}.
\end{align}

\noindent
Similarly, one can take covariant derivatives of $S_{GB}$. Our task is to use the Ladau-DeWitt gauge condition to simplify terms in \eqref{13}.

\begin{align}\label{B6}
    \Tilde{\Gamma}^k_{ij} \eta^i \eta^j &= \Gamma^k_{ij} \eta^i \eta^j + \dfrac{1}{2} \gamma^{\alpha\epsilon} \gamma^{\beta\sigma} K_{\alpha i} K_{\beta j} \Bigl( K^n_\epsilon K^k_{\sigma;n} + K^n_\sigma K^k_{\epsilon;n} \Bigr) \eta^i \eta^j - \gamma^{\alpha \beta} \Bigl( K_{\alpha i} K^k_{\beta;j} + \nonumber \\& \hspace{7mm} K_{\alpha j} K^k_{\beta;i} \Bigr)  \eta^i \eta^j \nonumber \\&
    = \Gamma^k_{ij} \eta^i \eta^j,
\end{align}

\noindent
in the Landau-DeWitt gauge $K_{\alpha i}[\varphi_*] \eta^i = 0$. Now we will prove some useful results which we will use later in this section.

\begin{equation}\label{B7}
         \sigma^i_{(2)} = - \dfrac{1}{2} \Tilde{\Gamma}^i_{jk} \eta^j \eta^k = - \dfrac{1}{2} \Gamma^i_{jk} \eta^j \eta^k
\end{equation}

\noindent
Proof of \eqref{B6} follows from \eqref{B7}. Now,

\begin{equation}\label{B8}
        \Tilde{\Gamma}^i_{jk,l} \eta^j \eta^k \eta^l = \Gamma^i_{jk,l} \eta^j \eta^k \eta^l
\end{equation}

\noindent
\textbf{Proof:} Take derivative of \eqref{B1},

\begin{align}\label{B9}
    T^i_{jk,l} \eta^j \eta^k \eta^l &= \bigg[(K^m_\sigma K^i_{\tau;m})_{,l} B^\sigma_{(j} B^\tau_{k)} + \dfrac{K^m_\sigma K^i_{\tau;m}}{2} (B^\sigma_{j,l} B^\tau_{k} + B^\sigma_{j} B^\tau_{k,l} + B^\sigma_{k,l} B^\tau_{j} + B^\sigma_{k} B^\tau_{j,l}) - \nonumber \\& \hspace{3mm} (K^i_{\alpha;j})_{,l} B^\alpha_k -  (K^i_{\alpha;k})_{,l} B^\alpha_j - K^i_{\alpha;j} B^\alpha_{k,l} - K^i_{\alpha;k} B^\alpha_{j,l} \bigg]\eta^j \eta^k \eta^l.
\end{align}

\noindent
We know that using the Landau-DeWitt gauge, $B^\alpha_i \eta^i = \gamma^{\alpha \beta} K_{i \beta} \eta^i = 0$

\begin{align}\label{B10}
    T^i_{jk,l} \eta^j \eta^k \eta^l &= - [K^i_{\alpha;j} B^\alpha_{k,l} + K^i_{\alpha;k} B^\alpha_{j,l}] \eta^j \eta^k \eta^l \nonumber \\& = - [K^i_{\alpha;j} \gamma^{\alpha \beta}_{,l} K_{k \beta} + K^i_{\alpha;j} \gamma^{\alpha \beta} K_{k \beta ,l} + K^i_{\alpha;k} \gamma^{\alpha \beta}_{,l} K_{j \beta} + K^i_{\alpha;k} \gamma^{\alpha \beta} K_{j \beta ,l}] \eta^j \eta^k \eta^l \nonumber \\& = - [K^i_{\alpha;j} \gamma^{\alpha \beta} K_{k \beta ,l} + K^i_{\alpha;k} \gamma^{\alpha \beta} K_{j \beta ,l}] \eta^j \eta^k \eta^l.
\end{align}

\noindent
Let's take the derivative of the Landau-DeWitt gauge,

\begin{align}\label{B11}
    &(K_{i \alpha} \eta^i)_{,j} = K_{i \alpha , j} \eta^i + K_{i \alpha} \delta^i_j = 0 \nonumber \\
    &K_{i \alpha , j} \eta^i = - K_{j \alpha}.
\end{align}

\noindent
Using \eqref{B11} in the above equation,

\begin{align}\label{B12}
    &T^i_{jk,l} \eta^j \eta^k \eta^l = K^i_{\alpha;j} \gamma^{\alpha \beta} K_{l \beta} \eta^j \eta^l + K^i_{\alpha;k} \gamma^{\alpha \beta} K_{l \beta} \eta^k \eta^l \nonumber \\
    &T^i_{jk,l} \eta^j \eta^k \eta^l = 0.
\end{align}

\noindent
Using \eqref{B12},

\begin{align}\label{B13}
    &\Tilde{\Gamma}^i_{jk,l} \eta^j \eta^k \eta^l = \Gamma^i_{jk,l} \eta^j \eta^k \eta^l + T^i_{jk,l} \eta^j \eta^k \eta^l \nonumber \\
    &\Tilde{\Gamma}^i_{jk,l} \eta^j \eta^k \eta^l = \Gamma^i_{jk,l} \eta^j \eta^k \eta^l.
\end{align}

\noindent
Now,

\begin{equation}\label{B14}
        \Tilde{\Gamma}^m_{ij,kl} \eta^i \eta^j \eta^k \eta^l = \Gamma^m_{ij,kl} \eta^i \eta^j \eta^k \eta^l.
\end{equation}

\noindent
\textbf{Proof:} Take derivative of \eqref{B9} and multiply $\eta$,

\begin{align*}
    &(T^m_{ij,k} \eta^i \eta^j \eta^k) _{,l} \eta^l = T^m_{ij,kl} \eta^i \eta^j \eta^k \eta^l + T^m_{ij,k} \eta^j \eta^k \delta^i_l \eta^l + T^m_{ij,k} \eta^i \eta^k \delta^j_l \eta^l + T^m_{ij,k} \eta^i \eta^j \delta^k_l \eta^l = 0 \\
    &T^m_{ij,kl} \eta^i \eta^j \eta^k \eta^l = 0.
\end{align*}

\noindent
Using the above result,

\begin{align}\label{B15}
    &\Tilde{\Gamma}^m_{ij,kl} \eta^i \eta^j \eta^k \eta^l = \Gamma^m_{ij,kl} \eta^i \eta^j \eta^k \eta^l + T^m_{ij,kl} \eta^i \eta^j \eta^k \eta^l \nonumber \\
    &\Tilde{\Gamma}^m_{ij,kl} \eta^i \eta^j \eta^k \eta^l = \Gamma^m_{ij,kl} \eta^i \eta^j \eta^k \eta^l.
\end{align}

\begin{equation}\label{B16}
    \sigma^m_{(3)} = - \dfrac{1}{3!} (\Tilde{\Gamma}^m_{ij,k} + \Tilde{\Gamma}^m_{in} \Tilde{\Gamma}^n_{jk} ) \eta^i \eta^j \eta^k = - \dfrac{1}{3!} (\Gamma^m_{ij,k} + \Tilde{\Gamma}^m_{in} \Gamma^n_{jk} ) \eta^i \eta^j \eta^k.
\end{equation}

\noindent
Proof of \eqref{B16} is pretty straightforward using \eqref{B6} and \eqref{B8}.

\noindent
Using the above results, we can calculate each term in \eqref{29}. These terms are given below,

\begin{align}\label{B17}
    &\dfrac{1}{4!} \Bar{S}_{;ijkl} \eta^i \eta^j \eta^k \eta^l = \dfrac{1}{4!} \Bigl[\Bar{S}_{,ijkl} - \Gamma^m_{ij} \Bar{S}_{,mkl} - \Gamma^m_{ik} S_{,mjl} - \Gamma^m_{il} \Bar{S}_{,mjk} - \Gamma^m_{jk} \Bar{S}_{,iml} - \Gamma^m_{jl} S_{,imk} - \nonumber \\& \hspace{35mm} \Gamma^m_{kl} S_{,ijm}   - \Gamma^m_{ij,k} \Bar{S}_{,ml} - \Gamma^m_{ij,l} \Bar{S}_{,mk} -  \Gamma^m_{ik,l} \Bar{S}_{,mj} - \Gamma^m_{jk,l} \Bar{S}_{,mi} + \Gamma^m_{ik} \Tilde{\Gamma}^n_{jm} \Bar{S}_{,nl}  \nonumber \\& \hspace{35mm} + \Gamma^m_{jk} \Tilde{\Gamma}^n_{im} \Bar{S}_{,nl} +  \Gamma^m_{il} \Tilde{\Gamma}^n_{jm} \Bar{S}_{,nk} + \Gamma^m_{il} \Tilde{\Gamma}^n_{km} \Bar{S}_{,nj} + \Gamma^m_{jl} \Tilde{\Gamma}^n_{im} \Bar{S}_{,nk} + \Gamma^m_{jl} \Tilde{\Gamma}^n_{mk} \Bar{S}_{,ni} \nonumber \\& \hspace{35mm} + \Gamma^m_{kl} \Tilde{\Gamma}^n_{im} \Bar{S}_{,nj} +  \Gamma^m_{kl} \Tilde{\Gamma}^n_{jm} \Bar{S}_{,ni} +  \Gamma^m_{kl} \Gamma^n_{ij} \Bar{S}_{,nm} + \Gamma^m_{jl} \Gamma^n_{ik} \Bar{S}_{,nm} + \Gamma^m_{ij} \Gamma^n_{kl} \Bar{S}_{,nm} \nonumber \\& \hspace{35mm} - \Gamma^m_{ij,kl} \Bar{S}_{,m} +  \Gamma^m_{ik,l} \Tilde{\Gamma}^n_{jm} \Bar{S}_{,n} + \Gamma^m_{jk,l} \Tilde{\Gamma}^n_{im} \Bar{S}_{,n} +  \Gamma^m_{ik} \Tilde{\Gamma}^n_{jm,l} \Bar{S}_{,n} + \Gamma^m_{jk} \Tilde{\Gamma}^n_{im,l} \Bar{S}_{,n}  \nonumber \\& \hspace{35mm} + \Gamma^m_{il} \Tilde{\Gamma}^n_{jm,k} \Bar{S}_{,n} + \Gamma^m_{jl} \Tilde{\Gamma}^n_{im,k} \Bar{S}_{,n} + \Gamma^m_{kl} \Tilde{\Gamma}^n_{ij,m} \Bar{S}_{,n} -  \Gamma^m_{il} \Tilde{\Gamma}^n_{mk} \Tilde{\Gamma}^p_{jn} \Bar{S}_{,p} \nonumber \\& \hspace{35mm} -  \Gamma^m_{jl} \Tilde{\Gamma}^n_{mk} \Tilde{\Gamma}^p_{in} \Bar{S}_{,p} -  \Gamma^m_{kl} \Tilde{\Gamma}^n_{im} \Tilde{\Gamma}^p_{jn} \Bar{S}_{,p} - \Gamma^m_{kl} \Tilde{\Gamma}^n_{jm} \Tilde{\Gamma}^p_{in} \Bar{S}_{,p} - \Gamma^m_{il} \Gamma^n_{jk} \Tilde{\Gamma}^p_{mn} \Bar{S}_{,p} - \nonumber \\& \hspace{35mm} \Gamma^m_{jl} \Gamma^n_{ik} \Tilde{\Gamma}^p_{mn} \Bar{S}_{,p} \Bigr] \eta^i \eta^j \eta^k \eta^l   \nonumber \\
    &\dfrac{1}{2} \Bar{S}_{;ijm} \eta^i \eta^j \sigma^m_{(2)} = - \dfrac{1}{4} \Bigl[\Bar{S}_{,ijm} \Gamma^m_{kl} - \Gamma^m_{kl} \Tilde{\Gamma}^n_{ij,m} \Bar{S}_{,n} - \Gamma^n_{ij} \Gamma^m_{kl} \Bar{S}_{,nm} - \Gamma^m_{kl} \Tilde{\Gamma}^n_{im} \Bar{S}_{,jn} - \Gamma^m_{kl} \Tilde{\Gamma}^n_{jm} \Bar{S}_{,in} + \nonumber \\& \hspace{35mm} \Gamma^m_{kl} \Tilde{\Gamma}^n_{im} \Tilde{\Gamma}^p_{jn} \Bar{S}_{,p} +  \Gamma^m_{kl} \Tilde{\Gamma}^n_{jm} \Tilde{\Gamma}^p_{in} \Bar{S}_{,p} \Bigr]\eta^i \eta^j \eta^k \eta^l  \nonumber \\ 
    &\Bar{S}_{;im} \eta^i \sigma^m_{(3)} = - \dfrac{1}{3!} \Bigl[\Bar{S}_{,im} \Gamma^m_{jk,l} + \Tilde{\Gamma}^m_{jn} \Gamma^n_{kl} \Bar{S}_{,im} - \Gamma^m_{jk,l} \Tilde{\Gamma}^n_{im} \Bar{S}_{,n} -  \Tilde{\Gamma}^m_{jn} \Tilde{\Gamma}^n_{kl} \Tilde{\Gamma}^p_{im} \Bar{S}_{,p} \Bigr]\eta^i \eta^j \eta^k \eta^l  \nonumber \\
    &\dfrac{1}{2} \Bar{S}_{;mn} \sigma^m_{(2)} \sigma^j_{(2)} = \dfrac{1}{8} \Bigl[\Gamma^m_{ij} \Gamma^n_{kl} \Bar{S}_{mn} - \Gamma^m_{ij} \Gamma^n_{kl} \Tilde{\Gamma}^p_{mn} \Bar{S}_{,p} \Bigr]\eta^i \eta^j \eta^k \eta^l  \nonumber \\ 
    &\dfrac{1}{3!} \Bar{S}_{;ijk} \eta^i \eta^j \eta^k = \dfrac{1}{3!} \Bigl[\Bar{S}_{,ijk} - \Gamma^m_{ij,k} \Bar{S}_{,m} - \Gamma^m_{ij} \Bar{S}_{,mk} - \Gamma^m_{ik} \Bar{S}_{,jm} - \Gamma^m_{jk} \Bar{S}_{,im} + \Gamma^m_{ik} \Tilde{\Gamma}^n_{jm} \Bar{S}_{,n} \nonumber \\& \hspace{35mm} + \Gamma^m_{jk} \Tilde{\Gamma}^n_{im} \Bar{S}_{,n} \Bigr]  \eta^i \eta^j \eta^k \nonumber \\
    &\Bar{S}_{im} \eta^i \sigma^m_{(2)} = - \dfrac{1}{2} \Bigl[\Bar{S}_{im} \Gamma^m_{jk} - \Gamma^m_{jk} \Tilde{\Gamma}^n_{im} \Bar{S}_{,n} \Bigr] \eta^i \eta^j \eta^k.
\end{align}

\noindent
Now we will calculate the covariant derivatives of $S_{GB}$ term. Before that, let's calculate the derivative of $S_{GB}$ with respect to the field. $S_{GB}$ is given by

\begin{equation}\label{B18}
    S_{GB} = \dfrac{f^{\alpha \beta}}{4 \ \alpha} \chi_\alpha \chi_\beta.
\end{equation}

\noindent
Where $f^{\alpha \beta}$ is a symmetric field-independent function. We take the derivative of \eqref{B18} in the background field limit.

\begin{align}\label{B19}
    &S_{GB,i} = \dfrac{f^{\alpha \beta}}{4 \ \alpha} (\chi_{\alpha,i} \chi_\beta + \chi_\alpha \chi_{\beta,i}) \nonumber \\
    &\Bar{S}_{GB,i} = \dfrac{f^{\alpha \beta}}{4 \ \alpha} (\Bar{\chi}_{\alpha,i} \Bar{\chi}_\beta + \Bar{\chi}_\alpha \Bar{\chi}_{\beta,i}).
\end{align}

\noindent
But,
\begin{align}\label{B20}
    &\chi_\alpha = g_{ij} K^i_\alpha (\phi^j - \Bar{\phi}^j) \nonumber \\
    &\Bar{\chi}_\alpha = g_{ij} \Bar{K}^i_\alpha (\Bar{\phi}^j - \Bar{\phi}^j) = 0.
\end{align}

\noindent
Using \eqref{B20} into \eqref{B19}, we get

\begin{equation}\label{B21}
    \Bar{S}_{GB,i} = 0.
\end{equation}

\noindent
Let's take 2nd derivative of $S_{GB}$,

\begin{align}\label{B22}
    &S_{GB,ij} = \dfrac{f^{\alpha \beta}}{4 \ \alpha} (\chi_{\alpha,ij} \chi_\beta + \chi_{\alpha,i} \chi_{\beta,j} + \chi_{\alpha,j} \chi_{\beta,i} + \chi_\alpha \chi_{\beta,ij})  \nonumber \\
    &\Bar{S}_{GB,ij} = \dfrac{f^{\alpha \beta}}{4 \ \alpha} (\Bar{\chi}_{\alpha,ij} \Bar{\chi}_\beta + \Bar{\chi}_{\alpha,i} \Bar{\chi}_{\beta,j} + \Bar{\chi}_{\alpha,j} \Bar{\chi}_{\beta,i} + \Bar{\chi}_\alpha \Bar{\chi}_{\beta,ij}) \nonumber \\
    &\Bar{S}_{GB,ij} = \dfrac{f^{\alpha \beta}}{4 \ \alpha} (\Bar{\chi}_{\alpha,i} \Bar{\chi}_{\beta,j} + \Bar{\chi}_{\alpha,j} \Bar{\chi}_{\beta,i}).
\end{align}

\noindent
It is easy to show that $S_{GB,ij\cdots} = 0$, because $\chi_{\alpha,ij\cdots} = 0$. Using \eqref{B21} and \eqref{B22} we can calculate terms involving $S_{GB}$ in \eqref{29}.

\begin{align}\label{B23}
    &\dfrac{1}{4!} \Bar{S}_{GB;ijkl} \eta^i \eta^j \eta^k \eta^l = \dfrac{1}{4!} \Bigl[ - \Gamma^m_{ij,k} \Bar{S}_{GB,ml} - \Gamma^m_{ij,l} \Bar{S}_{GB,mk} -  \Gamma^m_{ik,l} \Bar{S}_{GB,mj} - \Gamma^m_{jk,l} \Bar{S}_{GB,mi} + \nonumber \\& \hspace{35mm} \Gamma^m_{ik} \Tilde{\Gamma}^n_{jm}   \Bar{S}_{GB,nl} + \Gamma^m_{jk} \Tilde{\Gamma}^n_{im} \Bar{S}_{GB,nl} + \Gamma^m_{il} \Tilde{\Gamma}^n_{jm} \Bar{S}_{GB,nk} + \Gamma^m_{il} \Tilde{\Gamma}^n_{km} \Bar{S}_{GB,nj}  \nonumber \\& \hspace{35mm} + \Gamma^m_{jl} \Tilde{\Gamma}^n_{im} \Bar{S}_{GB,nk}   + \Gamma^m_{jl} \Tilde{\Gamma}^n_{mk} \Bar{S}_{GB,ni} + \Gamma^m_{kl} \Tilde{\Gamma}^n_{im} \Bar{S}_{GB,nj} + \Gamma^m_{kl} \Tilde{\Gamma}^n_{jm} \Bar{S}_{GB,ni} \nonumber \\& \hspace{35mm} + \Gamma^m_{kl} \Gamma^n_{ij} \Bar{S}_{GB,nm} +  \Gamma^m_{jl} \Gamma^n_{ik} \Bar{S}_{GB,nm}  + \Gamma^m_{ij} \Gamma^n_{kl} \Bar{S}_{GB,nm} \Bigr] \eta^i \eta^j \eta^k \eta^l \nonumber \\
    &\dfrac{1}{2} \Bar{S}_{GB;ijm} \eta^i \eta^j \sigma^m_{(2)} = - \dfrac{1}{4} \Bigl[- \Gamma^n_{ij} \Gamma^m_{kl} \Bar{S}_{GB,nm} - \Gamma^m_{kl} \Tilde{\Gamma}^n_{im} \Bar{S}_{GB,jn} - \Gamma^m_{kl} \Tilde{\Gamma}^n_{jm} \Bar{S}_{GB,in} \Bigr]\eta^i \eta^j \eta^k \eta^l \nonumber \\ 
    &\Bar{S}_{GB;im} \eta^i \sigma^m_{(3)} = - \dfrac{1}{3!} \Bigl[\Bar{S}_{GB,im} \Gamma^m_{jk,l} + \Tilde{\Gamma}^m_{jn} \Gamma^n_{kl} \Bar{S}_{GB,im} \Bigr]\eta^i \eta^j \eta^k \eta^l \nonumber \\
    &\dfrac{1}{2} \Bar{S}_{GB;mn} \sigma^m_{(2)} \sigma^j_{(2)} = \dfrac{1}{8} \Gamma^m_{ij} \Gamma^n_{kl} \Bar{S}_{GB,mn} \eta^i \eta^j \eta^k \eta^l \nonumber \\ 
    &\dfrac{1}{3!} \Bar{S}_{GB;ijk} \eta^i \eta^j \eta^k = \dfrac{1}{3!} \Bigl[  - \Gamma^m_{ij} \Bar{S}_{GB,mk} - \Gamma^m_{ik} \Bar{S}_{GB,jm} - \Gamma^m_{jk} \Bar{S}_{GB,im} \Bigr] \eta^i \eta^j \eta^k \nonumber \\
    &\Bar{S}_{GB;im} \eta^i \sigma^m_{(2)} = - \dfrac{1}{2} \Bar{S}_{GB,im} \Gamma^m_{jk} \eta^i \eta^j \eta^k.
\end{align}

\noindent
Now we will calculate terms involving $Q_{\alpha \beta}$ in \eqref{29}. Since all the higher order derivatives beyond the first order of $Q_{\alpha \beta}$ with respect to fields are zero, we find,

\begin{align}\label{B24}
    &\Bar{Q}_{\alpha \beta ;k} \Bar{c}^\alpha c^\beta \sigma^k_{(2)} = - \dfrac{1}{2} \Bar{Q}_{\alpha \beta ,k} \Gamma^k_{ij} \Bar{c}^\alpha c^\beta \eta^i \eta^j \nonumber \\
    &\dfrac{1}{2} \Bar{Q}_{\alpha \beta ;ij} \Bar{c}^\alpha c^\beta \eta^i \eta^j = - \dfrac{1}{2} \Bar{Q}_{\alpha \beta ,k} \Gamma^k_{ij} \Bar{c}^\alpha c^\beta \eta^i \eta^j \nonumber \\
    &\Bar{Q}_{\alpha \beta ;i} \Bar{c}^\alpha c^\beta \eta^i = \Bar{Q}_{\alpha \beta ,i} \Bar{c}^\alpha c^\beta \eta^i.
\end{align}

\section{Loop Integrals}\label{AppC}

\noindent
Most loop integrals are calculated using the well-known Passarino-Veltman integrals and tensor reduction \cite{bardin1999standard}.

\subsection{Integrals that appear in one-loop calculation}

\noindent
Integrals that appear in $\langle S_2 \rangle$ at one-loop in  Sec\ref{SecIIIA}.

\begin{flalign}\label{C1}
    &(1) \int \dfrac{d^4 k}{(2 \pi)^4} \dfrac{1}{k^2} = 0 \nonumber \\
    &(2) \int \dfrac{d^4 k}{(2 \pi)^4} \dfrac{1}{k^2 + m^2} = \dfrac{m^2}{16 \pi^2} \biggl[- \dfrac{2}{\epsilon} - 1 + \gamma_E + \text{ln}\biggl(\dfrac{m^2 }{4 \pi \mu^2}\biggr) \biggr]
\end{flalign}

\noindent
Integrals that appear in $\langle (S_1)^2 \rangle$ at one-loop in Sec\ref{SecIIIA}.

\begin{align}\label{C2}
    \int \dfrac{d^4 p}{(2 \pi)^4} \ \dfrac{1}{p^2((p-q)^2 + m^2)} = \dfrac{1}{16 \pi^2} \biggl[2 + \dfrac{2}{\epsilon} - \gamma_E - \text{ln}\biggl(\dfrac{m^2}{4 \pi \mu^2}\biggr) - \text{ln}\biggl(1 + \dfrac{q^2}{m^2}\biggr) \ \biggl(1 + \dfrac{m^2}{q^2}\biggr) \biggr]
\end{align}

\subsection{Integrals that appear in two-loops calculation}

\noindent
We start with loop integrals in \eqref{45}. (1), (2), and (3) of \eqref{45} vanish following arguments given in \cite{bardin1999standard}. Also, we will use the following results,

\begin{align}\label{C3}
    &\int \dfrac{d^4 k}{(2 \pi)^4} \dfrac{k^a}{k^2} = 0 \nonumber \\
    &\int \dfrac{d^4 k}{(2 \pi)^4} \dfrac{k^a}{k^4} = 0 \nonumber \\
    &\int \dfrac{d^4 k}{(2 \pi)^4} \dfrac{k^a}{k^6} = 0
\end{align}

\begin{align}\label{C4}
    \int \dfrac{d^4 p_1}{(2 \pi)^4} \dfrac{d^4 p_2}{(2 \pi)^4} \ \dfrac{1}{(p^2_1 + m^2) (p^2_2 + m^2)} &= \int \dfrac{d^4 p_1}{(2 \pi)^4} \  \dfrac{1}{p^2_1 + m^2} \ \int \dfrac{d^4 p_2}{(2 \pi)^4} \ \dfrac{1}{p^2_2 + m^2} \nonumber \\&  = \Biggl(\dfrac{m^2}{16 \pi^2} \biggl[- \dfrac{2}{\epsilon} - 1 + \gamma + \ln \biggl(\dfrac{m^2}{4 \pi \mu^2}\biggr) \biggr] \Biggl)^2 \nonumber \\& = \dfrac{m^4}{64 \pi^4 \epsilon^2} + \dfrac{m^4}{64 \pi^4 \epsilon} \Biggl(1 - \gamma_E - \ln \biggl(\dfrac{m^2}{4 \pi \mu^2}\biggr) \Biggr)
\end{align}

\begin{align}\label{C5}
    \int \dfrac{d^4 p_1}{(2 \pi)^4} \dfrac{d^4 p_2}{(2 \pi)^4} \ \dfrac{1}{p^{2n}_1 (p_1 + p_2)^2} &= \int \dfrac{d^4 p_1}{(2 \pi)^4} \ \dfrac{1}{p^{2n}_1} \ \int \dfrac{d^4 q}{(2 \pi)^4} \ \dfrac{1}{q^2} \nonumber \\& = 0
\end{align}

\noindent
In \eqref{C5} we used $p_1 + p_2 = q \Rightarrow p_2 = q - p_1$.

\begin{align}\label{C6}
    \int \dfrac{d^4 p_1}{(2 \pi)^4} \dfrac{d^4 p_2}{(2 \pi)^4} \ \dfrac{p^a_1 p_{2a}}{p^2_1 p^2_2 (p_1 + p_2)^2} &= \dfrac{1}{2} \int \dfrac{d^4 p_1}{(2 \pi)^4} \dfrac{d^4 p_2}{(2 \pi)^4} \Biggl[ \dfrac{1}{p^2_1 p^2_2} - \dfrac{1}{p^2_2 (p_1 + p_2)^2} - \dfrac{1}{p^2_1 (p_1 + p_2)^2}\Biggr] \nonumber \\& = 0 \hspace{1cm} (\text{Using} \hspace{2mm} 2 p^a_1 p_{2a} = (p_1 + p_2)^2 - p^2_1 - p^2_2 )
\end{align}

\begin{align}\label{C7}
    \int \dfrac{d^4 p_1}{(2 \pi)^4} \dfrac{d^4 p_2}{(2 \pi)^4} \ \dfrac{p^a_1 p^b_1 p_{2a} p_{2b}}{p^2_1 p^4_2 (p_1 + p_2)^2} &= \int \dfrac{d^4 p_1}{(2 \pi)^4} \dfrac{d^4 p_2}{(2 \pi)^4} \Biggl[ \dfrac{p^a_1 p_{2a}}{2 p^2_1 p^4_2} - \dfrac{p^a_1 p_{2a}}{2 p^4_2 (p_1 + p_2)^2} - \nonumber \\& \hspace{7mm} \dfrac{p^a_1 p_{2a}}{2 p^2_1 p^2_2(p_1 + p_2)^2}\Biggr] \nonumber \\& = - \int \dfrac{d^4 p_1}{(2 \pi)^4} \dfrac{d^4 p_2}{(2 \pi)^4} \dfrac{p^a_1 p_{2a}}{2 p^4_2 (p_1 + p_2)^2} \hspace{3mm} (\text{From} \hspace{2mm} \eqref{C3}, \eqref{C6}) \nonumber \\& =  \int \dfrac{d^4 q}{(2 \pi)^4} \dfrac{d^4 p_2}{(2 \pi)^4} \Biggl[ \dfrac{1}{4 q^2 p^2_2} - \dfrac{q^a p_{2a}}{4 p^4_2 q^2}\Biggr] \nonumber \\& = 0
\end{align}

\begin{align}\label{C8}
    \int \dfrac{d^4 p_1}{(2 \pi)^4} \dfrac{d^4 p_2}{(2 \pi)^4} \ \dfrac{1}{p^4_1 p^2_2 (p_1 + p_2)^2} &= 0 \hspace{1cm} (\text{Due to lack of physical scale})
\end{align}

\begin{align}\label{C9}
    \int \dfrac{d^4 p_1}{(2 \pi)^4} \dfrac{d^4 p_2}{(2 \pi)^4} \ \dfrac{p^a_1 p_{2a} }{p^4_1 p^2_2 (p_1 + p_2)^2} &= \int \dfrac{d^4 p_1}{(2 \pi)^4} \dfrac{d^4 p_2}{(2 \pi)^4} \Biggl[ \dfrac{1}{2 p^4_1 p^2_2} - \dfrac{1}{2 p^2_1 p^2_2 (p_1 + p_2)^2} - \dfrac{1}{2 p^4_1 (p_1 + p_2)^2}\Biggr] \nonumber \\& = 0 \hspace{5mm} (\text{From} \hspace{3mm} \eqref{C5}, \eqref{C8})
\end{align}

\begin{align}\label{C10}
    \int \dfrac{d^4 p_1}{(2 \pi)^4} \dfrac{d^4 p_2}{(2 \pi)^4} \ \dfrac{p^a_1 p^b_1 p_{2a} p_{2b}}{p^6_1 p^2_2 (p_1 + p_2)^2} &= \int \dfrac{d^4 p_1}{(2 \pi)^4} \dfrac{d^4 p_2}{(2 \pi)^4} \Biggl[ \dfrac{p^a_1 p_{2a}}{2 p^6_1 p^2_2} - \dfrac{p^a_1 p_{2a}}{2 p^4_1 p^2_2 (p_1 + p_2)^2} - \dfrac{p^a_1 p_{2a}}{2 p^6_1 (p_1 + p_2)^2}\Biggr] \nonumber \\& = \int \dfrac{d^4 p_1}{(2 \pi)^4} \dfrac{d^4 q}{(2 \pi)^4} \Biggl[ \dfrac{1}{4 p^2_1 q^2} - \dfrac{p^a_1 q_a}{4 p^4_1 q^2} \Biggr] \hspace{5mm} (\text{From} \eqref{C3}, \eqref{C9}) \nonumber \\& = 0 \hspace{1cm} (\text{Using} \hspace{2mm} p_2 = q - p_1)
\end{align}

\begin{align}\label{C11}
    \int \dfrac{d^4 p_1}{(2 \pi)^4} \dfrac{d^4 p_2}{(2 \pi)^4} \ \dfrac{p^a_1 p^b_1 p^c_1 p_{2a} p_{2b} p_{2c}}{p^4_1 p^4_2 (p_1 + p_2)^2} &= \int \dfrac{d^4 p_1}{(2 \pi)^4} \dfrac{d^4 p_2}{(2 \pi)^4} \Biggl[ \dfrac{p^a_1 p^b_1 p_{2a} p_{2b}}{2 p^4_1 p^4_2} - \dfrac{p^a_1 p^b_1 p_{2a} p_{2b}}{2 p^4_1 p^2_2 (p_1 + p_2)^2} - \nonumber \\& \hspace{7mm} \dfrac{p^a_1 p^b_1 p_{2a} p_{2b}}{2 p^4_1 p^2_2 (p_1 + p_2)^2}\Biggr] \nonumber \\& = 0 \hspace{1cm} (\text{From} \hspace{2mm} \eqref{C7})
\end{align}

\begin{align}\label{C12}
    \int \dfrac{d^4 p_1}{(2 \pi)^4} \dfrac{d^4 p_2}{(2 \pi)^4} \ \dfrac{p^a_1 p_{2a}}{(p^2_1 + m^2) (p^2_2 + m^2)} &= 0 \hspace{1cm} (\text{Odd function of} \ p_1, p_2)
\end{align}

\begin{align}\label{C13}
    \int \dfrac{d^4 p_1}{(2 \pi)^4} \dfrac{d^4 p_2}{(2 \pi)^4} \ \dfrac{p^2_1}{p^2_2 (p_1 + p_2)} &= \int \dfrac{d^4 p_2}{(2 \pi)^4} \dfrac{d^4 q}{(2 \pi)^4} \Biggl[ \dfrac{1}{p^2_2} - \dfrac{2 p^a_2 q_a}{p^2_2 q^2} + \dfrac{1}{q^2} \Biggr] \hspace{1cm} (\text{Using} \hspace{2mm} p_1 = q - p_2) \nonumber \\& = 0 
\end{align}

\begin{align}\label{C14}
    \int \dfrac{d^4 p_1}{(2 \pi)^4} \dfrac{d^4 p_2}{(2 \pi)^4} \ \dfrac{p^a_1 p_{2a}}{p^2_1 (p_1 + p_2)^2} &= \int \dfrac{d^4 p_1}{(2 \pi)^4} \dfrac{d^4 p_2}{(2 \pi)^4} \Biggl[ \dfrac{1}{2 p^2_1} - \dfrac{1}{2 (p_1 + p_2)^2} - \dfrac{p^2_2}{p^2_1 (p_1 + p_2)^2} \Biggr] \nonumber \\& = 0 \hspace{1cm} (\text{From} \hspace{2mm} \eqref{C1},\eqref{C13})
\end{align}

\begin{align}\label{C15}
    \int \dfrac{d^4 p_1}{(2 \pi)^4} \dfrac{d^4 p_2}{(2 \pi)^4} \ \dfrac{p^a_1 p^b_1 p_{2a} p_{2b}}{p^2_1 p^2_2 (p_1 + p_2)^2} &= \int \dfrac{d^4 p_1}{(2 \pi)^4} \dfrac{d^4 p_2}{(2 \pi)^4} \Biggl[ \dfrac{1}{2 p^2_1 p^2_2} - \dfrac{1}{2 p^2_2 (p_1 + p_2)^2} - \dfrac{1}{p^2_1 (p_1 + p_2)^2} \Biggr] \nonumber \\& = 0 \hspace{1cm} (\text{From} \hspace{2mm} \eqref{C3},\eqref{C14})
\end{align}

\begin{align}\label{C16}
    \int \dfrac{d^4 p_1}{(2 \pi)^4} \dfrac{d^4 p_2}{(2 \pi)^4} \ \dfrac{p^2_1}{(p^2_1 + m^2) (p^2_2 + m^2)} &= \int \dfrac{d^4 p_1}{(2 \pi)^4} \dfrac{d^4 p_2}{(2 \pi)^4} \ \Biggl[ \dfrac{1}{p^2_2 + m^2} - \dfrac{m^2}{(p^2_1 + m^2) (p^2_2 + m^2)} \Biggr] \nonumber \\& = - \dfrac{m^6}{64 \pi^4 \epsilon^2} - \dfrac{m^6}{64 \pi^4 \epsilon} \Biggl(1 - \gamma_E - \ln \biggl(\dfrac{m^2}{4 \pi \mu^2}\biggr) \Biggr)
\end{align}

\noindent
For (11) and (12) loop integrals of \eqref{45}, we will use the following results from \cite{kleinert2001critical},

\begin{align}\label{C17}
     \int \dfrac{d^4 p_1}{(2 \pi)^4} \dfrac{d^4 p_2}{(2 \pi)^4} \ \dfrac{1}{p^2_1 (p^2_2 + m^2) (p_1 + p_2)^2} &= - \dfrac{1}{(4 \pi)^4} \Biggl[\dfrac{2 m^2}{\epsilon^2} + \dfrac{1}{\epsilon} \Biggl(m^2 (3 - 2 \gamma_E) - \nonumber \\& \hspace{7mm} 2 m^2 \ln \biggl(\dfrac{m^2}{4 \pi \mu^2}\biggr) \Biggr) \Biggr]
\end{align}

\begin{align}\label{C18}
     \int \dfrac{d^4 p_1}{(2 \pi)^4} \dfrac{d^4 p_2}{(2 \pi)^4} \ \dfrac{1}{p^4_1 (p^2_2 + m^2) (p_1 + p_2)^2} &= - \dfrac{1}{(4 \pi)^4} \Biggl[\dfrac{2}{\epsilon^2} + \dfrac{1}{\epsilon} \Biggl((1 - 2 \gamma_E) - 2 \ln \biggl(\dfrac{m^2}{4 \pi \mu^2}\biggr) \Biggr) \Biggr]
\end{align}

\begin{align}\label{C19}
    \int \dfrac{d^4 p_1}{(2 \pi)^4} \dfrac{d^4 p_2}{(2 \pi)^4} \ \dfrac{p^a_1 p_{2a}}{p^2_1 (p^2_2 + m^2) (p_1 + p_2)^2} &= \int \dfrac{d^4 p_1}{(2 \pi)^4} \dfrac{d^4 p_2}{(2 \pi)^4} \Biggl[ \dfrac{1}{2 p^2_1 (p^2_2 + m^2)} - \nonumber \\& \hspace{-3cm} \dfrac{1}{2 (p^2_2 + m^2) (p_1 + p_2)^2} +  \dfrac{m^2}{2 p^2_1 (p^2_2 + m^2) (p_1 + p_2)^2} \Biggr] \nonumber \\& = \int \dfrac{d^4 p_1}{(2 \pi)^4} \dfrac{d^4 p_2}{(2 \pi)^4} \ \dfrac{m^2}{2 p^2_1 (p^2_2 + m^2) (p_1 + p_2)^2} \nonumber \\& = - \dfrac{1}{2 (4 \pi)^4} \Biggl[\dfrac{2 m^4}{\epsilon^2} + \dfrac{1}{\epsilon} \Biggl(m^4 (3 - 2 \gamma_E) - \nonumber \\& \hspace{7mm} 2 m^4 \ln \biggl(\dfrac{m^2}{4 \pi \mu^2}\biggr) \Biggr) \Biggr]
\end{align}

\begin{align}\label{C20}
    \int \dfrac{d^4 p_1}{(2 \pi)^4} \dfrac{d^4 p_2}{(2 \pi)^4} \ \dfrac{p^a_1 p^b_1 p_{2a} p_{2b}}{p^4_1 (p^2_2 + m^2) (p_1 + p_2)^2} &= \int \dfrac{d^4 p_1}{(2 \pi)^4} \dfrac{d^4 p_2}{(2 \pi)^4} \Biggl[ \dfrac{p^a_1 p_{2a}}{2 p^4_1 (p^2_2 + m^2)} - \nonumber \\& \hspace{-3cm} \dfrac{p^a_1 p_{2a}}{2 p^2_1 (p^2_2 + m^2) (p_1 + p_2)^2} +  \dfrac{p^a_1 p_{2a} m^2}{2 p^4_1 (p^2_2 + m^2) (p_1 + p_2)^2} \Biggr] \nonumber \\& = \int \dfrac{d^4 p_1}{(2 \pi)^4} \dfrac{d^4 p_2}{(2 \pi)^4} \ \Biggl[\dfrac{m^2}{4 p^4_1 (p^2_2 + m^2)} - \nonumber \\& \hspace{-3cm} \dfrac{m^2}{2 p^2_1 (p^2_2 + m^2) (p_1 + p_2)^2}  + \dfrac{m^4}{4 p^4_1 (p^2_2 + m^2) (p_1 + P_2)^2} \Biggr] \nonumber \\& = - \dfrac{m^4}{512 \pi^4 \epsilon^2} + \dfrac{3 m^4}{1024 \pi^4 \epsilon}
\end{align}

\noindent
Using the \eqref{C4}, \eqref{C16}, \eqref{C19}, and \eqref{C20}, divergent part of loop integrals in \eqref{45} are given by,

\begin{align}\label{C21}
    &(4) \int \dfrac{d^4 p_1}{(2 \pi)^4} \dfrac{d^4 p_2}{(2 \pi)^4} \ \dfrac{1}{(p^2_1 + m^2) (p^2_2 + m^2)} = \dfrac{m^4}{64 \pi^4 \epsilon^2} + \dfrac{m^4}{64 \pi^4 \epsilon} \Biggl(1 - \gamma_E - \ln \biggl(\dfrac{m^2}{4 \pi \mu^2}\biggr) \Biggr) \nonumber \\
    &(11) \int \dfrac{d^4 p_1}{(2 \pi)^4} \dfrac{d^4 p_2}{(2 \pi)^4} \ \dfrac{p^a_1 p^b_1 p_{2a} p_{2b}}{p^4_1 (p^2_2 + m^2) (p_1 + p_2)^2} = - \dfrac{m^4}{512 \pi^4 \epsilon^2} + \dfrac{3 m^4}{1024 \pi^4 \epsilon} \nonumber \\
    &(12) \int \dfrac{d^4 p_1}{(2 \pi)^4} \dfrac{d^4 p_2}{(2 \pi)^4} \ \dfrac{p^a_1 p_{2a} }{p^2_1 (p^2_2 + m^2) (p_1 + p_2)^2} = - \dfrac{1}{2 (4 \pi)^4} \Biggl[\dfrac{2 m^4}{\epsilon^2} + \dfrac{1}{\epsilon} \Biggl(m^4 (3 - 2 \gamma_E) - \nonumber \\& \hspace{8cm} 2 m^4 \ln \biggl(\dfrac{m^2}{4 \pi \mu^2}\biggr) \Biggr) \Biggr] \nonumber \\
    &(14) \int \dfrac{d^4 p_1}{(2 \pi)^4} \dfrac{d^4 p_2}{(2 \pi)^4} \ \dfrac{p^2_1}{(p^2_1 + m^2) (p^2_2 + m^2)} =  - \dfrac{m^6}{64 \pi^4 \epsilon^2} - \dfrac{m^6}{64 \pi^4 \epsilon} \Biggl(1 - \gamma_E - \ln \biggl(\dfrac{m^2}{4 \pi \mu^2}\biggr) \Biggr)
\end{align}

\noindent
Apart from (4), (11), (12), and (14), all other loop integrals vanish.

\begin{align}\label{C22}
    (21) \int \dfrac{d^4 p_1}{(2 \pi)^4} \dfrac{d^4 p_2}{(2 \pi)^4} \ \dfrac{p^a_1 p^b_1 p^c_1 p_{2a} p_{2b} p_{2c}}{p^4_1 p^4_2} &= \int \dfrac{d^4 p_1}{(2 \pi)^4} \ \dfrac{p^a_1 p^b_1 p^c_1}{p^4_1} \ \int \dfrac{d^4 p_2}{(2 \pi)^4} \ \dfrac{p_{2a} p_{2b} p_{2c}}{p^4_2} \nonumber \\& = 0 \hspace{1cm} (\text{Odd function of} \ p_1, p_2 )
\end{align}

\noindent
We calculate loop integrals in \eqref{56}. The result of (1), (2), and (3) loop integrals are in \cite{kleinert2001critical} and \cite{tomsbook}. Please keep in mind that we will not write terms proportional to the function of $q$ as argued in the \eqref{55} in Sec.\ref{SecIIIC}.

\begin{align}\label{C23}
    \int \dfrac{d^4 p_1}{(2 \pi)^4} \dfrac{d^4 p_2}{(2 \pi)^4} \ \dfrac{1}{p^2_1 p^2_2 ((q - p_1 - p_2)^2 + m^2)} &= - \dfrac{1}{(4 \pi)^4} \Biggl[\dfrac{2 m^2}{\epsilon^2} + \dfrac{1}{\epsilon} \Biggl(m^2 (3 - 2 \gamma_E) - \nonumber \\& \hspace{7mm} 2 m^2 \ln \biggl(\dfrac{m^2}{4 \pi \mu^2}\biggr) \Biggr) \Biggr]
\end{align}

\begin{align}\label{C24}
    \int \dfrac{d^4 p_1}{(2 \pi)^4} \dfrac{d^4 p_2}{(2 \pi)^4} \ \dfrac{1}{p^2_1 (p^2_2 + m^2) ((q - p_1 - p_2)^2 + m^2)} &= - \dfrac{1}{(4 \pi)^4} \Biggl[\dfrac{4 m^2}{\epsilon^2} + \dfrac{1}{\epsilon} \Biggl(2 m^2 (3 - 2 \gamma_E) - \nonumber \\& \hspace{5mm} 4 m^2 \ln \biggl(\dfrac{m^2}{4 \pi \mu^2}\biggr) \Biggr) \Biggr]
\end{align}

\begin{align}\label{C25}
    \int \dfrac{d^4 p_1}{(2 \pi)^4} \dfrac{d^4 p_2}{(2 \pi)^4} \ \dfrac{1}{(p^2_1 + m^2) (p^2_2 + m^2) ((q - p_1 - p_2)^2 + m^2)} &= \dfrac{-1}{(4 \pi)^4} \Biggl[\dfrac{6 m^2}{\epsilon^2} + \nonumber \\& \hspace{-3cm} \dfrac{1}{\epsilon} \Biggl(3 m^2 (3 - 2 \gamma_E)  - 6 m^2 \ln \biggl(\dfrac{m^2}{4 \pi \mu^2}\biggr) \Biggr) \Biggr]
\end{align}

\begin{align}\label{B26}
    \int \dfrac{d^4 p_1}{(2 \pi)^4} \dfrac{d^4 p_2}{(2 \pi)^4} \ \dfrac{p^a_1 p^b_1 p_{2a} p_{2b}}{p^2_1 p^4_2} &= \int \dfrac{d^4 p_1}{(2 \pi)^4} \ \dfrac{p^a_1 p^b_1}{p^2_1} \ \int \dfrac{d^4 p_2}{(2 \pi)^4} \ \dfrac{p_{2a} p_{2b}}{p^4_2} \nonumber \\& = 0
\end{align}

\begin{align}\label{C27}
    \int \dfrac{d^4 p_1}{(2 \pi)^4} \dfrac{d^4 p_2}{(2 \pi)^4} \ \dfrac{1}{p^2_1} &= \int \dfrac{d^4 p_1}{(2 \pi)^4} \ \dfrac{1}{p^2_1} \ \int \dfrac{d^4 p_2}{(2 \pi)^4} \nonumber \\& = 0
\end{align}

\begin{align}\label{C28}
    \int \dfrac{d^4 p_1}{(2 \pi)^4} \dfrac{d^4 p_2}{(2 \pi)^4} \ \dfrac{p^a_1 p^b_1 p_{2a} p_{2b}}{p^4_1 p^4_2 ((q - p_1 - p_2)^2 + m^2)} &= \int \dfrac{d^4 p_1}{(2 \pi)^4} \dfrac{d^4 p_2}{(2 \pi)^4} \Biggl[ \dfrac{p^a_1 p_{2a} (p_1 + p_2)^2}{2 p^4_1 p^4_2 ((q - p_1 - p_2)^2 + m^2)} \nonumber \\& \hspace{-3cm} - \dfrac{p^a_1 p_{2a}}{2 p^2_1 p^4_2 ((q - p_1 - p_2)^2 + m^2)} -  \dfrac{p^a_1 p_{2a}}{2 p^4_1 p^2_2 ((q - p_1 - p_2)^2 + m^2)} \Biggr] \nonumber \\& = \int \dfrac{d^4 p_1}{(2 \pi)^4} \dfrac{d^4 p_2}{(2 \pi)^4} \ \Biggl[\dfrac{-m^2 p^a_1 p_{2a}}{2 p^4_1 p^4_2 ((q - p_1 - p_2)^2 + m^2)} \nonumber \\& \hspace{-3cm} - \dfrac{p^a_1 p_{2a}}{2 p^2_1 p^4_2 ((q - p_1 - p_2)^2 + m^2)}  + \dfrac{p^a_1 p_{2a}}{2 p^4_1 p^2_2 ((q - p_1 - p_2)^2 + m^2)} \Biggr]
\end{align}

\noindent
In the second step of \eqref{C28} we used $(p_1 + p_2)^2 = (q - (p_1 + p_2))^2 - q^2 + 2 q^a (p_1 + p_2)_a$. We ignored the term proportional to $q$, as argued before. Again using $p^a_1 p_{2a} = \dfrac{(p_1 + p_2)^2 - p^2_1 - p^2_2}{2}$ and $(p_1 + p_2)^2 = (q - (p_1 + p_2))^2 - q^2 + 2 q^a (p_1 + p_2)_a$ in the second step of \eqref{C28} to get,

\begin{align}\label{C29}
    \int \dfrac{d^4 p_1}{(2 \pi)^4} \dfrac{d^4 p_2}{(2 \pi)^4} \ \dfrac{p^a_1 p^b_1 p_{2a} p_{2b}}{p^4_1 p^4_2 ((q - p_1 - p_2)^2 + m^2)} &= - \dfrac{1}{4} \int \dfrac{d^4 p_1}{(2 \pi)^4} \dfrac{d^4 p_2}{(2 \pi)^4} \Biggl[ \dfrac{m^2}{p^4_1 p^4_2} -  \nonumber \\& \hspace{-7cm} \dfrac{2}{p^4_2 ((q - p_1 - p_2)^2 + m^2)} - \dfrac{2}{p^2_1 p^2_2 ((q - p_1 - p_2)^2 + m^2)} -  \dfrac{4}{p^2_1 p^4_2 ((q - p_1 - p_2)^2 + m^2)} \Biggr] \nonumber \\& = \dfrac{1}{128 \pi^4 \epsilon^2} - \dfrac{m^2}{64 \pi^4 \epsilon^2} + \dfrac{1}{256 \pi^4 \epsilon} + \dfrac{m^2}{512 \pi^4 \epsilon} \nonumber \\& \hspace{7mm} - \dfrac{\gamma_E}{128 \pi^4 \epsilon} -  \dfrac{\ln \biggl(\dfrac{m^2}{4 \pi \mu^2}\biggr)}{128 \pi^4 \epsilon}
\end{align}

\begin{equation}\label{C30}
     \int \dfrac{d^4 p_1}{(2 \pi)^4} \dfrac{d^4 p_2}{(2 \pi)^4} \ \dfrac{1}{p^4_2 ((q - p_1 - p_2)^2 + m^2)} \equiv \int \dfrac{d^4 p_1}{(2 \pi)^4} \dfrac{d^4 p_2}{(2 \pi)^4} \ \dfrac{1}{p^4_2 (p^2_1 + m^2)}
\end{equation}

\noindent
While deriving \eqref{C30}, we used $l_a = q_a - p_{1a} - p_{2a} \Rightarrow p_{1a} = l_a - q_a + p_{2a}$ keeping $q$ and $p_2$ constant. After that we rebel $l_a$ to $p_{1a}$

\begin{align}\label{C31}
    \int \dfrac{d^4 p_1}{(2 \pi)^4} \dfrac{d^4 p_2}{(2 \pi)^4} \ \dfrac{p^a_1 p^b_1 (q - p_1 - p_2)_a (q - p_1 - p_2)_b}{p^2_1 (p^2_1 + m^2) p^2_2 (q - p_1 - p_2)^4} &= \int \dfrac{d^4 p_1}{(2 \pi)^4} \dfrac{d^4 p_2}{(2 \pi)^4} \ \dfrac{p^a_1 p^b_1 }{p^2_1 (p^2_1 + m^2) } \nonumber \\& \hspace{7mm} \times \dfrac{(p_1 + p_2)_a (p_1 + p_2)_b}{p^2_2 (q - p_1 - p_2)^4} \nonumber \\& = \int \dfrac{d^4 p_1}{(2 \pi)^4} \dfrac{d^4 p_2}{(2 \pi)^4} \times \nonumber \\& \hspace{-3cm} \Biggl[\dfrac{p^2_1 + m^2 - m^2}{(p^2_1 + m^2) p^2_2 (q - p_1 - p_2)^4}  + \dfrac{2 p^a_1 p_{2a}}{(p^2_1 + m^2) p^2_2 (q - p_1 - p_2)^4} \nonumber \\& \hspace{7mm} + \dfrac{p^a_1 p_{2a} p^b_1 p_{2b}}{p^2_1 (p^2_1 + m^2) p^2_2 (q - p_1 - p_2)^4} \Biggr]
\end{align}

\noindent
From the first step to the second step in \eqref{C30}, we ignored terms proportional to $q$. Using $p^a_1 p_{2a} = \dfrac{(p_1 + p_2)^2 - p^2_1 - p^2_2}{2}$ and $(p_1 + p_2)^2 = (q - (p_1 + p_2))^2 - q^2 + 2 q^a (p_1 + p_2)_a$ in the second step of \eqref{C31} to get,

\begin{align}\label{C32}
    \int \dfrac{d^4 p_1}{(2 \pi)^4} \dfrac{d^4 p_2}{(2 \pi)^4} \ \dfrac{p^a_1 p^b_1 (q - p_1 - p_2)_a (q - p_1 - p_2)_b}{p^2_1 (p^2_1 + m^2) p^2_2 (q - p_1 - p_2)^4} &= \dfrac{1}{4} \int \dfrac{d^4 p_1}{(2 \pi)^4} \dfrac{d^4 p_2}{(2 \pi)^4} \Biggl[ \dfrac{2}{(p^2_1 + m^2) p^2_2 } \times \nonumber \\& \hspace{-5.5cm} \dfrac{1}{(q - p_1 - p_2)^2} - \dfrac{m^2}{(p^2_1 + m^2) p^2_2 (q - p_1 - p_2)^4} - \dfrac{1}{(p^2_1 + m^2) (q - p_1 - p_2)^4} \Biggr] \nonumber \\& = - \dfrac{m^2}{512 \pi^4 \epsilon^2} + \dfrac{7 m^2}{1024 \pi^4 \epsilon} - \dfrac{m^2 \gamma_E}{256 \pi^4 \epsilon} - \nonumber \\& \hspace{7mm} \dfrac{m^2 \ln \biggl(\dfrac{m^2}{4 \pi \mu^2}\biggr)}{128 \pi^4 \epsilon}
\end{align}

\begin{align}\label{C33}
    \int \dfrac{d^4 p_1}{(2 \pi)^4} \dfrac{d^4 p_2}{(2 \pi)^4} \ \dfrac{p^a_1 p^b_1 p_{2a} p^c_2 (q - p_1 - p_2)_b (q - p_1 - p_2)_c}{p^2_1 (p^2_1 + m^2) p^4_2 (q - p_1 - p_2)^4} &= \int \dfrac{d^4 p_1}{(2 \pi)^4} \dfrac{d^4 p_2}{(2 \pi)^4} \dfrac{p^a_1 p^b_1 p_{2a} p^c_2 }{p^2_1 (p^2_1 + m^2) } \nonumber \\& \hspace{7mm} \times \dfrac{(p_1 + p_2)_b (p_1 + p_2)_c}{p^4_2 (q - p_1 - p_2)^4} \nonumber \\& = \int \dfrac{d^4 p_1}{(2 \pi)^4} \dfrac{d^4 p_2}{(2 \pi)^4} \ \Biggl[\dfrac{p^a_1 p_{2a} }{(p^2_1 + m^2) } \times \nonumber \\& \hspace{-7cm} \dfrac{p^b_1 p_{2b}}{p^4_2 (q - p_1 - p_2)^4} + \dfrac{p^a_1 p_{2a}}{(p^2_1 + m^2) p^2_2 (q - p_1 - p_2)^4} +  \dfrac{p^a_1 p_{2a} p^b_1 p_{2b} p^c_1 p_{2c}}{p^2_1 (p^2_1 + m^2) p^4_2 (q - p_1 - p_2)^4} \nonumber \\& \hspace{7mm} + \dfrac{p^a_1 p_{2a} p^b_1 p_{2b}}{p^2_1 (p^2_1 + m^2) p^2_2 (q - p_1 - p_2)^4} \Biggr]
\end{align}

\noindent
Using $p^a_1 p_{2a} = \dfrac{(p_1 + p_2)^2 - p^2_1 - p^2_2}{2}$ and $(p_1 + p_2)^2 = (q - (p_1 + p_2))^2 - q^2 + 2 q^a (p_1 + p_2)_a$ in the second step of \eqref{C33} to get,

\begin{align}\label{C34}
    \int \dfrac{d^4 p_1}{(2 \pi)^4} \dfrac{d^4 p_2}{(2 \pi)^4} \ \dfrac{p^a_1 p^b_1 p_{2a} p^c_2 (q - p_1 - p_2)_b (q - p_1 - p_2)_c}{p^2_1 (p^2_1 + m^2) p^4_2 (q - p_1 - p_2)^4} &= - \dfrac{1}{8} \int \dfrac{d^4 p_1}{(2 \pi)^4} \dfrac{d^4 p_2}{(2 \pi)^4} \Biggl[ \dfrac{2 m^2}{p^4_1 p^4_2} - \nonumber \\& \hspace{-5cm} \dfrac{4}{(p^2_1 + m^2) p^2_2 (q - p_1 - p_2)^2} - \dfrac{m^2}{(p^2_1 + m^2) p^2_2 (q - p_1 - p_2)^4} \Biggr] \nonumber \\& = - \dfrac{7 m^2}{1024 \pi^4 \epsilon^2} + \dfrac{13 m^2}{2048 \pi^4 \epsilon} - \dfrac{5 m^2 \gamma_E}{1024 \pi^4 \epsilon}  \nonumber \\& \hspace{7mm} - \dfrac{5 m^2 \ln \biggl(\dfrac{m^2}{4 \pi \mu^2}\biggr)}{1024 \pi^4 \epsilon}
\end{align}

\begin{align}\label{C35}
    \int \dfrac{d^4 p_1}{(2 \pi)^4} \dfrac{d^4 p_2}{(2 \pi)^4} \ \dfrac{p^a_1 p^b_1 p^c_1 p^d_1 p_{2a} p_{2b} (q - p_1 - p_2)_c (q - p_1 - p_2)_d}{p^4_1 (p^2_1 + m^2) p^4_2 (q - p_1 - p_2)^4} & \nonumber \\& \hspace{-5cm} = \int \dfrac{d^4 p_1}{(2 \pi)^4} \dfrac{d^4 p_2}{(2 \pi)^4} \ \dfrac{p^a_1 p^b_1 p^c_1 p^d_1 p_{2a} p_{2b} (p_1 + p_2)_c (p_1 + p_2)_d}{p^4_1 (p^2_1 + m^2) p^4_2 (q - p_1 - p_2)^4} \nonumber \\&\hspace{-5cm} = \int \dfrac{d^4 p_1}{(2 \pi)^4} \dfrac{d^4 p_2}{(2 \pi)^4} \ \Biggl[\dfrac{p^a_1 p_{2a} p^b_1 p_{2b}}{(p^2_1 + m^2) p^4_2 (q - p_1 - p_2)^4} \nonumber \\& \hspace{-7cm} + \dfrac{2 p^a_1 p_{2a} p^b_1 p_{2b} p^c_1 p_{2c}}{p^2_1 (p^2_1 + m^2) p^4_2 (q - p_1 - p_2)^4} +  \dfrac{p^a_1 p_{2a} p^b_1 p_{2b} p^c_1 p_{2c} p^d_1 p_{2d}}{p^2_1 (p^2_1 + m^2) p^4_2 (q - p_1 - p_2)^4} \Biggr]
\end{align}

\noindent
Using $p^a_1 p_{2a} = \dfrac{(p_1 + p_2)^2 - p^2_1 - p^2_2}{2}$ and $(p_1 + p_2)^2 = (q - (p_1 + p_2))^2 - q^2 + 2 q^a (p_1 + p_2)_a$ in the second step of \eqref{C35} to get,

\begin{align}\label{C36}
    \int \dfrac{d^4 p_1}{(2 \pi)^4} \dfrac{d^4 p_2}{(2 \pi)^4} \ \dfrac{p^a_1 p^b_1 p^c_1 p^d_1 p_{2a} p_{2b} (q - p_1 - p_2)_c (q - p_1 - p_2)_d}{p^4_1 (p^2_1 + m^2) p^4_2 (q - p_1 - p_2)^4} &= - \dfrac{1}{16} \int \dfrac{d^4 p_1}{(2 \pi)^4} \dfrac{d^4 p_2}{(2 \pi)^4} \Biggl[ \dfrac{m^2}{p^4_1 p^4_2} + \nonumber \\& \hspace{-7cm} \dfrac{4}{(p^2_1 + m^2) p^4_2}  - \dfrac{14}{(p^2_1 + m^2) p^2_2 (q - p_1 - p_2)^2}  + \dfrac{5 m^2}{(p^2_1 + m^2) p^2_2 (q - p_1 - p_2)^4} \Biggr] \nonumber \\& = - \dfrac{13 m^2}{2048 \pi^4 \epsilon^2} + \dfrac{45 m^2}{4096 \pi^4 \epsilon} - \nonumber \\& \hspace{-1cm} \dfrac{13 m^2 \gamma_E}{2048 \pi^4 \epsilon} - \dfrac{13 m^2 \ln \biggl(\dfrac{m^2}{4 \pi \mu^2}\biggr)}{2048 \pi^4 \epsilon}
\end{align}

\begin{align}\label{C37}
    \int \dfrac{d^4 p_1}{(2 \pi)^4} \dfrac{d^4 p_2}{(2 \pi)^4} \ \dfrac{(q - p_1 - p_2)^2}{(p^2_1 + m^2) (p^2_2 + m^2)} &= \int \dfrac{d^4 p_1}{(2 \pi)^4} \dfrac{d^4 p_2}{(2 \pi)^4} \Biggl[\dfrac{p^2_1 + m^2 - m^2}{(p^2_1 + m^2) (p^2_2 + m^2)} + \nonumber \\& \hspace{7mm} \dfrac{p^2_2 + m^2 - m^2}{(p^2_1 + m^2) (p^2_2 + m^2)}  + \dfrac{2 p^a_1 p_{2a}}{(p^2_1 + m^2) (p^2_2 + m^2)} \Biggr] \nonumber \\& = -2 \int \dfrac{d^4 p_1}{(2 \pi)^4} \dfrac{d^4 p_2}{(2 \pi)^4} \dfrac{m^2}{(p^2_1 + m^2) (p^2_2 + m^2)} \nonumber \\& = - \dfrac{m^6}{32 \pi^4 \epsilon^2} - \dfrac{m^6}{32 \pi^4 \epsilon} \Biggl(1 - \gamma_E - \ln \biggl(\dfrac{m^2}{4 \pi \mu^2}\biggr) \Biggr)
\end{align}

\begin{align}\label{C38}
    \int \dfrac{d^4 p_1}{(2 \pi)^4} \dfrac{d^4 p_2}{(2 \pi)^4} \ \dfrac{p^a_1 p^b_1 p_{2a} p_{2b}(q - p_1 - p_2)^2}{p^4_1 p^4_2} &= \int \dfrac{d^4 p_1}{(2 \pi)^4} \dfrac{d^4 p_2}{(2 \pi)^4} \ \dfrac{p^a_1 p^b_1 p_{2a} p_{2b}(p_1 + p_2)^2}{p^4_1 p^4_2} \nonumber \\& =  \int \dfrac{d^4 p_1}{(2 \pi)^4} \dfrac{d^4 p_2}{(2 \pi)^4} \ \Biggl[\dfrac{p^a_1 p^b_1 p_{2a} p_{2b}}{p^2_1 p^4_2} + \nonumber \\& \hspace{7mm} \dfrac{p^a_1 p^b_1 p_{2a} p_{2b}}{p^4_1 p^2_2} + \dfrac{2 p^a_1 p^b_1 p^c_1 p_{2a} p_{2b} p_{2c}}{p^4_1 p^4_2} \Biggr] \nonumber \\&= 0
\end{align}

\begin{align}\label{C39}
    \int \dfrac{d^4 p_1}{(2 \pi)^4} \dfrac{d^4 p_2}{(2 \pi)^4} \ \dfrac{(q - p_1 - p_2)^a (q - p_1 - p_2)^b p_{2a} p_{2b}}{p^2_1 p^4_2} &= \int \dfrac{d^4 p_1}{(2 \pi)^4} \dfrac{d^4 p_2}{(2 \pi)^4} \ \dfrac{p_{2a} p_{2b}}{p^2_1} \times \nonumber \\& \hspace{7mm} \dfrac{(p_1 + p_2)^a (p_1 + p_2)^b}{ p^4_2} \nonumber \\& =  \int \dfrac{d^4 p_1}{(2 \pi)^4} \dfrac{d^4 p_2}{(2 \pi)^4} \ \Biggl[\dfrac{p^a_1 p^b_1 p_{2a} p_{2b}}{p^2_1 p^4_2} \nonumber \\& \hspace{7mm} + \dfrac{2 p^a_1 p_{2a}}{p^2_1 p^2_2} + \dfrac{1}{p^2_1} \Biggr] \nonumber \\&= 0
\end{align}

\begin{align}\label{C40}
    \int \dfrac{d^4 p_1}{(2 \pi)^4} \dfrac{d^4 p_2}{(2 \pi)^4} \ \dfrac{(q - p_1 - p_2)^2}{p^2_1 p^2_2} &= \int \dfrac{d^4 p_1}{(2 \pi)^4} \dfrac{d^4 p_2}{(2 \pi)^4} \ \dfrac{(p_1 + p_2)^2}{p^2_1 p^2_2} \nonumber \\& =  \int \dfrac{d^4 p_1}{(2 \pi)^4} \dfrac{d^4 p_2}{(2 \pi)^4} \ \Biggl[\dfrac{1}{p^2_1} + \dfrac{1}{p^2_1} + \dfrac{2 p^a_1 p_{2a}}{p^2_1 p^2_2} \Biggr] \nonumber \\&= 0
\end{align}

\begin{align}\label{C41}
    \int \dfrac{d^4 p_1}{(2 \pi)^4} \dfrac{d^4 p_2}{(2 \pi)^4} \ \dfrac{p^a_1 p^b_1 p_{2a} p_{2b}}{p^2_1 p^2_2 (q - p_1 - p_2)^2} &= \int \dfrac{d^4 p_1}{(2 \pi)^4} \dfrac{d^4 p_2}{(2 \pi)^4} \Biggl[\dfrac{p^a_1 p_{2a}}{2 p^2_1 p^2_2} - \dfrac{1}{4 p^2_2} + \dfrac{p^2_1}{4 p^2_2 (q - p_1 - p_2)^2}  \nonumber \\& \hspace{-1.1cm} + \dfrac{1}{(q - p_1 - p_2)^2} - \dfrac{1}{4 p^2_1} + \dfrac{p^2_2}{4 p^2_1 (q - p_1 - p_2)^2} + \dfrac{1}{(q - p_1 - p_2)^2} \Biggr]  \nonumber \\&= 0
\end{align}

\noindent
We have used $p^a_1 p_{2a} = \dfrac{(p_1 + p_2)^2 - p^2_1 - p^2_2}{2}$ and $(p_1 + p_2)^2 = (q - (p_1 + p_2))^2 - q^2 + 2 q^a (p_1 + p_2)_a$ in \eqref{C41}.

\begin{align}\label{C42}
    \int \dfrac{d^4 p_1}{(2 \pi)^4} \dfrac{d^4 p_2}{(2 \pi)^4} \ \dfrac{p^a_1 p^b_1 p_{2a} p^b_2 (q - p_1 - p_2)_b (q - p_1 - p_2)_c}{p^4_1 p^2_2 (q - p_1 - p_2)^2} &= \int \dfrac{d^4 p_1}{(2 \pi)^4} \dfrac{d^4 p_2}{(2 \pi)^4} \ \dfrac{p^a_1 p^b_1 p_{2a} p^b_2 }{p^4_1 p^2_2 } \nonumber \\& \hspace{7mm} \times \dfrac{(p_1 + p_2)_b (p_1 + p_2)_c}{(q - p_1 - p_2)^2}  \nonumber \\&=  \int \dfrac{d^4 p_1}{(2 \pi)^4} \dfrac{d^4 p_2}{(2 \pi)^4} \Biggl[\dfrac{p^a_1 p^b_1 p_{2a} p_{2b}}{p^2_1 p^2_2 (q - p_1 - p_2)^2}  \nonumber \\& \hspace{-5cm} + \dfrac{p^a_1 p_{2a}}{p^2_1 (q - p_1 - p_2)^2} + \dfrac{p^a_1 p^b_1 p^c_1 p_{2a} p_{2b} p_{2c}}{p^4_1 p^2_2 (q - p_1 - p_2)^2} +  \dfrac{p^a_1 p^b_1 p_{2a} p_{2b}}{p^4_1 (q - p_1 - p_2)^2}\Biggr]
\end{align}

\noindent
Using $p^a_1 p_{2a} = \dfrac{(p_1 + p_2)^2 - p^2_1 - p^2_2}{2}$, $(p_1 + p_2)^2 = (q - (p_1 + p_2))^2 - q^2 + 2 q^a (p_1 + p_2)_a$, and \eqref{C41} in \eqref{C42}, we see that \eqref{C42} vanish.

\begin{align}\label{C43}
    \int \dfrac{d^4 p_1}{(2 \pi)^4} \dfrac{d^4 p_2}{(2 \pi)^4} \ \dfrac{p^a_1 p^b_1 p_{2a} p^b_2 (q - p_1 - p_2)_b (q - p_1 - p_2)_c}{p^4_1 p^2_2 (q - p_1 - p_2)^2} &= 0
\end{align}

\noindent
Using the \eqref{C23}, \eqref{C24}, \eqref{C25}, \eqref{C29}, \eqref{C32}, \eqref{C34}, \eqref{C36} and \eqref{C37}, divergent part of loop integrals in \eqref{56} are given by,

\begin{align}\label{C44}
    &(1) \int \dfrac{d^4 p_1}{(2 \pi)^4} \dfrac{d^4 p_2}{(2 \pi)^4} \ \dfrac{1}{p^2_1 p^2_2 ((q - p_1 - p_2)^2 + m^2)} = - \dfrac{1}{(4 \pi)^4} \Biggl[\dfrac{2 m^2}{\epsilon^2} + \dfrac{1}{\epsilon} \Biggl(m^2 (3 - 2 \gamma_E) - \nonumber \\& \hspace{8.2cm} 2 m^2 \ln \biggl(\dfrac{m^2}{4 \pi \mu^2}\biggr) \Biggr) \Biggr]  \nonumber \\
    &(2) \int \dfrac{d^4 p_1}{(2 \pi)^4} \dfrac{d^4 p_2}{(2 \pi)^4} \ \dfrac{1}{p^2_1 (p^2_2 + m^2) ((q - p_1 - p_2)^2 + m^2)} = - \dfrac{1}{(4 \pi)^4} \Biggl[\dfrac{4 m^2}{\epsilon^2} + \nonumber \\& \hspace{8cm} \dfrac{1}{\epsilon} \Biggl(2 m^2 (3 - 2 \gamma_E)   - 4 m^2 \ln \biggl(\dfrac{m^2}{4 \pi \mu^2}\biggr) \Biggr) \Biggr]  \nonumber \\
    &(3) \int \dfrac{d^4 p_1}{(2 \pi)^4} \dfrac{d^4 p_2}{(2 \pi)^4} \ \dfrac{1}{(p^2_1 + m^2) (p^2_2 + m^2) ((q - p_1 - p_2)^2 + m^2)} = - \dfrac{1}{(4 \pi)^4} \Biggl[\dfrac{6 m^2}{\epsilon^2} + \nonumber \\& \hspace{8cm} \dfrac{1}{\epsilon} \Biggl(3 m^2 (3 - 2 \gamma_E)  - 6 m^2 \ln \biggl(\dfrac{m^2}{4 \pi \mu^2}\biggr) \Biggr) \Biggr] \nonumber \\
    &(4) \int \dfrac{d^4 p_1}{(2 \pi)^4} \dfrac{d^4 p_2}{(2 \pi)^4} \ \dfrac{p^a_1 p^b_1 p_{2a} p_{2b}}{p^4_1 p^4_2 ((p_1 + p_2 - q)^2 + m^2)} = \dfrac{1}{128 \pi^4 \epsilon^2} - \dfrac{m^2}{64 \pi^4 \epsilon^2} + \dfrac{1}{256 \pi^4 \epsilon} + \dfrac{m^2}{512 \pi^4 \epsilon} \nonumber \\& \hspace{8.2cm} - \dfrac{\gamma_E}{128 \pi^4 \epsilon} -  \dfrac{\ln \biggl(\dfrac{m^2}{4 \pi \mu^2}\biggr)}{128 \pi^4 \epsilon} \nonumber \\
    &(5) \int \dfrac{d^4 p_1}{(2 \pi)^4} \dfrac{d^4 p_2}{(2 \pi)^4} \ \dfrac{p^a_1 p^b_1 (q - p_1 - p_2)_{a} (q - p_1 - p_2)_{b}}{p^2_1 (p^2_1 + m^2) p^2_2 (p_1 + p_2 - q)^4} = - \dfrac{m^2}{512 \pi^4 \epsilon^2} + \dfrac{7 m^2}{1024 \pi^4 \epsilon} - \dfrac{m^2 \gamma_E}{256 \pi^4 \epsilon} - \nonumber \\& \hspace{9.2cm} \dfrac{m^2 \ln \biggl(\dfrac{m^2}{4 \pi \mu^2}\biggr)}{128 \pi^4 \epsilon} \nonumber \\
    &(6) \int \dfrac{d^4 p_1}{(2 \pi)^4} \dfrac{d^4 p_2}{(2 \pi)^4} \ \dfrac{p^a_1 p^b_1 p_{2a} p^c_2 (q - p_1 - p_2)_{b} (q - p_1 - p_2)_{c}}{p^2_1 (p^2_1 + m^2) p^4_2 (p_1 + p_2 - q)^4} = - \dfrac{7 m^2}{1024 \pi^4 \epsilon^2} + \dfrac{13 m^2}{2048 \pi^4 \epsilon} - \nonumber \\& \hspace{10cm} \dfrac{5 m^2 \gamma_E}{1024 \pi^4 \epsilon} -  \dfrac{5 m^2 \ln \biggl(\dfrac{m^2}{4 \pi \mu^2}\biggr)}{1024 \pi^4 \epsilon} \nonumber \\
    &(7) \int \dfrac{d^4 p_1}{(2 \pi)^4} \dfrac{d^4 p_2}{(2 \pi)^4} \ \dfrac{p^a_1 p^b_1 p^c_1 p^d_1 p_{2a} p_{2b} (q - p_1 - p_2)_{c} (q - p_1 - p_2)_{d}}{p^4_1 (p^2_1 + m^2) p^4_2 (p_1 + p_2 - q)^4} = - \dfrac{13 m^2}{2048 \pi^4 \epsilon^2} + \dfrac{45 m^2}{4096 \pi^4 \epsilon} - \nonumber \\& \hspace{9.5cm} \dfrac{13 m^2 \gamma_E}{2048 \pi^4 \epsilon}  - \dfrac{13 m^2 \ln \biggl(\dfrac{m^2}{4 \pi \mu^2}\biggr)}{2048 \pi^4 \epsilon} \nonumber \\
    &(8) \int \dfrac{d^4 p_1}{(2 \pi)^4} \dfrac{d^4 p_2}{(2 \pi)^4} \ \dfrac{(q - p_1 - p_2)^2}{(p^2_1 + m^2) (p^2_2 + m^2)} = - \dfrac{m^6}{32 \pi^4 \epsilon^2} - \dfrac{m^6}{32 \pi^4 \epsilon} \Biggl(1 - \gamma_E - \ln \biggl(\dfrac{m^2}{4 \pi \mu^2}\biggr) \Biggr)
\end{align}


 \bibliographystyle{JHEP}
 \bibliography{biblio.bib}

\end{document}